# Non-Unique Machine Learning Mapping in Data-Driven Reynolds Averaged Turbulence Models

Anthony Man, Mohammad Jadidi, Amir Keshmiri, Hujun Yin, Yasser Mahmoudi[*]

School of Engineering, The University of Manchester, Manchester, M13 9PL, UK

*Corresponding author: yasser.mahmoudilarimi@manchester.ac.uk

## Abstract

Recent growing interest in using machine learning for turbulence modelling has led to many proposed data-driven turbulence models in the literature. However, most of these models have not been developed with overcoming non-unique mapping (NUM) in mind, which is a significant source of training and prediction error. Only NUM caused by one-dimensional channel flow data has been well studied in the literature, despite most data-driven models having been trained on two-dimensional flow data. The present work aims to be the first detailed investigation on NUM caused by two-dimensional flows. A method for quantifying NUM is proposed and demonstrated on data from a flow over periodic hills, and an impinging jet. The former is a wall-bounded separated flow, and the latter is a shear flow containing stagnation and recirculation. This work confirms that data from two-dimensional flows can cause NUM in data-driven turbulence models with the commonly used invariant inputs. This finding was verified with both cases, which contain different flow phenomena, hence showing that NUM is not limited to specific flow physics. Furthermore, the proposed method revealed that regions containing low strain and rotation or near pure shear cause the majority of NUM in both cases – approximately 76% and 89% in the flow over periodic hills and impinging jet, respectively. These results led to viscosity ratio being selected as a supplementary input variable (SIV), demonstrating that SIVs can reduce NUM caused by data from two-dimensional flows and subsequently improve the accuracy of tensor-basis machine learning models for turbulence modelling.

## Nomenclature

| Variable | Meaning | Unit |
|---|---|---|
| $b_{ij}$ | Anisotropy tensor | - |
| $B$ | Impinging jet inlet width | m |
| $C_\mu$ | Boussinesq hypothesis parameter | - |
| $g_n$ | Scalar coefficients of the general effective-viscosity hypothesis | - |
| $d$ | Distance between two arbitrary scatter points $p$ and $q$ | - |
| $H_h$ | Hill height | m |
| $k$ | Turbulent kinetic energy | $m^2/s^2$ |
| $L_x$ | Domain length in periodic hills case | m |
| $n_{CI}$ | Number of conflicting instances in a grid cell | - |
| $r_\nu$ | Viscosity ratio ($r_\nu = \nu_t/(100\nu + \nu_t)$) | - |
| $\boldsymbol{R}$ | Non-dimensional mean rotation rate ($\boldsymbol{R} = kr_{ij}/\varepsilon$) | - |
| $r_{ij}$ | Mean rotation rate | 1/s |
| $Re_t$ | Turbulent Reynolds number ($Re_t = k^2/\nu\varepsilon$) | - |
| $\boldsymbol{S}$ | Non-dimensional mean strain rate ($\boldsymbol{S} = ks_{ij}/\varepsilon$) | - |
| $s_{ij}$ | Mean strain rate | 1/s |
| $u^+$ | Nondimensional streamwise velocity ($u^+ = \bar{u}_1/u_\tau$) | - |
| $u_\tau$ | Friction velocity ($\sqrt{\tau_w/\rho}$) | m/s |
| $u_i'$ | Velocity fluctuation in i$^{th}$ direction | m/s |
| $\bar{u}_i$ | Mean velocity in i$^{th}$ direction | m/s |
| $U_b$ | Bulk inlet velocity in periodic hills case | m/s |
| $U_{in}$ | Uniform inlet velocity in impinging jet case | m/s |
| $x_i$ | Distance in i$^{th}$ direction | m |
| $\boldsymbol{x}_{tr}$ | Input variables of the two-dimensional GEVH ($\boldsymbol{x}_{tr} \equiv \{tr(\boldsymbol{S}^2), tr(\boldsymbol{R}^2)\}$) | - |
| $y^+$ | Nondimensional distance from the wall ($y^+ = x_2 u_\tau/\nu$) | - |

| Symbol | | |
|---|---|---|
| $\tau_w$ | Wall shear stress | $m^2/s^2$ |
| $\tau_{ij}$ | Reynolds stress | $m^2/s^2$ |
| $\delta_{ij}$ | Kronecker delta | - |
| $\varepsilon$ | Turbulent kinetic energy dissipation rate | $m^2/s^3$ |
| $\alpha$ | Non-dimensional shear velocity gradient ($\alpha = (k/\varepsilon)(\partial \bar{u}_1/\partial x_2)$) | - |
| $\nu$ | Kinematic molecular viscosity | $m^2/s$ |
| $\nu_t$ | Kinematic eddy viscosity | $m^2/s$ |
| $\rho$ | Density | kg/m³ |
| $\beta$ | Steepness factor in periodic hills case | - |

| Subscript | | |
|---|---|---|
| 1 | Streamwise direction | |
| 2 | Wall-normal direction | |
| 3 | Spanwise direction | |
| $p$ | Arbitrary scatter point $p$ | |
| $q$ | Arbitrary scatter point $q$ | |

| **Superscript** | |
|---|---|
| $\overline{\square}$ | Reynolds-averaged |
| $\square$ | Min-max normalized quantity |

| **Abbreviation** | |
|---|---|
| CFD | Computational fluid dynamics |
| CLA | Complete linkage agglomerative |
| CI | Conflicting instance |
| DNS | Direct numerical simulation |
| GEVH | General effective-viscosity hypothesis |
| LES | Large eddy simulation |
| LSR | Low strain and rotation |
| ML | Machine learning |
| NNM | Nearest neighbours method |
| NUM | Non-unique mapping |
| PBM | Proximity box method |
| RANS | Reynolds-averaged Navier-Stokes |
| SIV | Supplementary input variable |
| SRCC | Spearman's rank correlation coefficient |
| SST | Shear stress transport |
| TBML | Tensor-basis machine learning |
| TBNN | Tensor-basis neural network |
| TKE | Turbulent kinetic energy |

# 1   Introduction

Over the last few decades, advancement in computational power and algorithms has greatly improved the speed and capability of computational fluid dynamics (CFD) solvers for simulating turbulent flows. The continual drive for better accuracy and wider applicability has led to the development of a range of approaches including Reynolds-averaged Navier-Stokes (RANS) models, and scale-resolving methods such as large eddy simulation (LES) and direct numerical simulation (DNS) (Slotnick *et al.* 2014). However, scale-resolving methods are still currently computationally infeasible for most industrial CFD practitioners (Durbin 2018). Using RANS turbulence models is still therefore the most common approach for simulating turbulent flows, and it is projected that this will remain for the near future (Duraisamy *et al.* 2017; Bush *et al.* 2019). Despite their popularity however, progress in improving RANS model accuracy has stalled, due to the challenge of conserving their high robustness and low computational cost in the development of new models (Xiao & Cinnella 2019). As a result, RANS approaches still give inaccurate predictions for many complex flows, including flow separation, and impingement (Duraisamy *et al.* 2017).

Improvements in computational resources and recent advances in machine learning (ML) methodologies have also led to rapid developments in ML, resulting in its prevalence in many technological applications today including natural language processing, and computer vision (Goodfellow *et al.* 2016; Gupta *et al.* 2021). More recently, there has been growing interest in using data-driven ML models as an alternative to existing RANS turbulence models or to augment them due to the stagnation in RANS development. Yarlanki *et al.* (2012) explored the capabilities of neural networks for calibrating RANS model coefficients. Similarly, Parish and Duraisamy (2016) demonstrated that ML can be used to tune the magnitude of terms in the RANS equations. Although improved predictions given by both models were reported, their approaches still adopted a linear relationship between mean strain rate and Reynolds stress, which neglects rotation and higher order strain rate terms (Singh *et al.* 2017). Complex flow features that RANS models fail to predict accurately such as high streamline curvature, secondary vortices and anisotropic turbulence are subsequently not fully accounted for or at all in these approaches (Lien *et al.* 1996). Therefore, Reynolds stress (or its projections) should be selected as the ML model outputs.

Modelling Reynolds stress using a tensor basis has emerged as a popular data-driven approach (Duraisamy *et al.* 2019). These models – referred to as tensor-basis machine learning (TBML) models in this paper – are based on predicting the scalar coefficients of a Galilean-invariant tensor integrity basis, which gives a closed-form expression for Reynolds stress in homogeneous turbulent flows (Pope 1975). Ling *et al.* (2016) developed the first TBML model called the tensor-basis neural network (TBNN), which reportedly predicted Reynolds stress anisotropy and mean flow fields more accurately than a linear and nonlinear RANS model when tested on canonical flow cases. As the tensor basis sensitises the Reynolds stress to mean strain rate, mean rotation rate and their higher order products, these effects are fully accounted for in the TBNN. Various enhancements to the TBNN have thereafter been proposed in the literature. Zhang *et al.* (2019) introduced regularisation in the loss function and included $y^+$ as an input variable. Parmar *et al.* (2020) expanded the tensor basis to include the effect of mean pressure gradient. Jiang *et al.* (2021) proposed an additional ML model for predicting the tensor basis and a different timescale for non-dimensionalising mean strain and rotation rate. Some studies have reported improved prediction accuracy after adopting more advanced ML techniques, including ensemble learning by Man *et al.* (2022) and modular TBNNs by Man *et al.* (2023). The latter involves using multiple TBNNs – each trained and tested on specific regions of flow physics. Modifying the TBNN has also been investigated, *e.g.*, with skipped connections by Jiang *et al.* (2021) and with Bayesian inference by Tang *et al.* (2023). Different ML model types have also been explored for predicting the scalars, including symbolic regression (Weatheritt & Sandberg 2016; 2017), random forests (Kaandorp & Dwight 2020), and gradient-boosted tree models (McConkey *et al.* 2022).

While this literature review shows that many TBML models have been proposed, most have not been developed with overcoming non-unique mapping (NUM) in mind (also known as the multi-value problem). This is a well-known source of prediction error that can occur when a deterministic ML model is trained on data containing one-to-many relations – *i.e.*, where multiple observations in the training dataset have the same input values but different output values (Bishop 2006). This ill-posed problem of conflicting observations can cause difficulty in training because their same input values lead to the loss function being minimised against different target output values, thus posing a challenge in achieving model convergence (Liu *et al.* 2018). Although some investigations in the literature have explored NUM in TBML models (Liu *et al.* 2021; Jiang *et al.* 2021; Cai *et al.* 2022), only NUM resulting from one-dimensional (1D) channel flow data has been well studied, and their proposed solutions, which involved modifying or including supplementary input variables were developed to eliminate NUM due to near-wall flow data exclusively. Given that most TBML models in the literature to date have been trained on data from two-dimensional (2D) flow cases containing various flow physics instead, it is crucial to investigate whether NUM in TBML models can be caused by 2D flow data. If so, an awareness in the community to develop steps that reduce or eliminate NUM resulting from data of such flows becomes necessary to improve the training convergence and subsequent predictive accuracy of these models.

Although extensive literature on modelling one-to-many relations exists in mathematics and other engineering disciplines, limited studies were found on methods for identifying them. Intuitively, one-to-many relations may be identified by creating hypersurfaces of the outputs in the input space, and detecting if the hypersurfaces overlap in the input space (Shizawa 1994). The data may alternatively be clustered into groups that correspond to these hypersurfaces and assessed whether the clusters share the same input space, which was attempted by Huang *et al.* (2013). However, both methods do not quantify the extent of one-to-many relations in the data, which would be helpful in analysing the worst affected regions in a chosen flow case. Various ML models have been proposed in the literature that can represent one-to-many relations between inputs and outputs which would address NUM, including recurrent neural networks (Uno *et al.* 1995), mixture density networks (Bishop 2006), and mixture of experts (Jacobs *et al.* 1992). However, as these approaches would introduce significant complexity to existing TBML models, their exploration was reserved for future work. It was considered reasonable instead to attempt extending the use of a supplementary input variable (SIV) seen in 1D channel flow studies to 2D flows for this investigation.

Overall, the current work aims to investigate non-unique mapping (NUM) in data-driven turbulence models based on the tensor-basis approach for two-dimensional (2D) flows. Although true anisotropy has normally been used as the output to identify NUM resulting from 1D channel flow data, we show that this is unsuitable for 2D flows, as this can lead to false diagnoses of NUM in the tensor basis. The true scalar coefficients that tensor-basis machine learning (TBML) models aim to predict are proposed as the output for identifying NUM instead. A clustering approach for identifying and quantifying NUM throughout the input space before TBML model training is proposed. Based on unsupervised learning, this approach does not require user action, and can be used with any number of inputs. The approach was demonstrated using data from two different flow scenarios: a wall-bounded separated flow over periodic hills case, and a free shear flow in an impinging jet case. It is shown that NUM can exist in TBML models trained on these 2D flows with the standard invariant inputs. Moreover, this finding is verified with data from two very different flow cases, showing that NUM is not limited to certain flow physics in the training data. The method was extended by including a supplementary input variable (SIV) and repeated with the two cases, which demonstrated that SIVs can reduce NUM from 2D flow cases in TBML models and – with verification from a training experiment, – subsequently improve TBML model accuracy.

## 2 Background

### 2.1 General Effective Viscosity Hypothesis

The general effective-viscosity hypothesis (GEVH) postulated by Pope (1975) states that the Reynolds stress $\tau_{ij}$ under local equilibrium in Reynolds-averaged Navier Stokes (RANS) modelling of a turbulent flow can be fully determined by a finite sum of basis tensors (Zhou *et al.* 2021). For two-dimensional flows where the velocity and variation of mean quantities are negligible or zero in one co-ordinate direction, the GEVH contains four terms (Pope 1975):

$$\tau_{ij} = \overline{u_i' u_j'} = 2k\left(b_{ij} + \frac{1}{3}\delta_{ij}\right) \tag{2.1}$$

where, $k$ is the turbulent kinetic energy (TKE) and $\delta_{ij}$ is Kronecker delta. The anisotropy tensor $b_{ij}$ in Eq. (2.1) is a normalised form of the Reynolds stress and contains three of the terms:

$$b_{ij} = g_1 \mathbf{S} + g_2 (\mathbf{SR} - \mathbf{RS}) + g_3 \left(\mathbf{S}^2 - \frac{1}{3} tr(\mathbf{S}^2)\delta_{ij}\right) \tag{2.2}$$

$\mathbf{S}\left(= k s_{ij}/\varepsilon\right)$ and $\mathbf{R}\left(= k r_{ij}/\varepsilon\right)$ are the mean strain and mean rotation rate, respectively, non-dimensionalised by TKE and TKE dissipation rate $\varepsilon$. The dimensional mean strain rate $s_{ij}$ and mean rotation rate $r_{ij}$ can be calculated as:

$$s_{ij} = \frac{1}{2}\left(\frac{\partial \bar{u}_i}{\partial x_j} + \frac{\partial \bar{u}_j}{\partial x_i}\right), \qquad r_{ij} = \frac{1}{2}\left(\frac{\partial \bar{u}_i}{\partial x_j} - \frac{\partial \bar{u}_j}{\partial x_i}\right) \tag{2.3}$$

Lowercase $s$ and $r$ are used to denote the dimensionalised form of the mean strain and rotation rate tensors, while uppercase $S$ and $R$ are used to denote the non-dimensionalised form. The tensor products $\mathbf{SR}$, $\mathbf{RS}$ and $\mathbf{S}^2$ in Eq. (2.2) can be rewritten in Einstein notation, *e.g.*, $\mathbf{SR} = k^2 s_{ik} r_{kj}/\varepsilon^2$. The notation $tr(...)$ denotes the trace of the product inside the parentheses. These traces are also known as invariants and can be represented in Einstein notation, *e.g.*, $tr(\mathbf{S}^2) = k^2 s_{ik} s_{ki}/\varepsilon^2$. Coefficients $g_1$, $g_2$ and $g_3$ are scalars that are unknown functions of the following invariants of $\mathbf{S}$ and $\mathbf{R}$:

$$g_n = f\left(tr(\mathbf{S}^2), tr(\mathbf{R}^2)\right), \qquad (n = 1, 2, 3) \tag{2.4}$$

It has been shown in the literature that Eq. (2.2) can be represented by a TBML model (Ling *et al.* 2016). Given invariants such as those in Eq. (2.4) as inputs, these models are trained to predict the $g_n$ coefficients that when combined with $\mathbf{S}$ and $\mathbf{R}$ from a RANS method on the right-hand side of Eq. (2.2), the summation gives a $b_{ij}$ result equivalent to the true $b_{ij}$ from experiments or an accurate high-fidelity simulation. These $g_n$ targets will be referred to as true $g_n$.

The first TBML model was the TBNN introduced by Ling *et al.* (2016). This was demonstrated with the three-dimensional version of Eq. (2.2), which contains five input invariants of $\boldsymbol{S}$ and $\boldsymbol{R}$, and ten output $g_n$ coefficients (n = 1 to 10) – corresponding to ten basis tensors. Ling *et al.* (2016) showed that the TBNN was able to learn the relationship between these inputs from RANS and true $b_{ij}$ from LES and DNS for some classical flow cases during training (Kutz 2017). This capability has now also been demonstrated with other TBML models, *e.g.*, the tensor-basis random forest by Kaandorp and Dwight (2020). Although most TBML models were developed to be applicable to three-dimensional flows, they can also be used to model $b_{ij}$ in one- and two-dimensional flows. In these situations, the velocity and its mean gradients would be zero in one or more co-ordinate directions, resulting in some negligible and redundant terms. More specifically for two-dimensional flows, the TBML models effectively become a representation of Eq. (2.2) with inputs given in Eq. (2.4), as the other seven tensors become zero or can be subsumed into the three tensors in Eq. (2.2), and the other three inputs become zero or can be represented by the inputs in Eq. (2.4). In particular, the quadratic term $g_4(\boldsymbol{R}^2 - tr(\boldsymbol{R}^2)\delta_{ij}/3)$ that exists in the full GEVH can be subsumed into the $g_3$ term in Eq. (2.2). For 1D flows where only shear velocity gradient $\partial \bar{u}_1/\partial x_2$ is the only non-negligible velocity gradient component, the GEVH and subsequent TBML model simplifies to modelling $b_{ij}$ as a function of $\partial \bar{u}_1/\partial x_2$. The simplicity of the one-dimensional GEVH has made it an ideal starting point for demonstrating NUM in TBML models (Liu *et al.* 2021).

## 2.2 Non-Unique Mapping

Non-unique mapping (NUM) can be defined as a poor fitting of a function on a dataset, due to an occurrence or occurrences of multiple solutions satisfying a particular array of input variables in some observations of the dataset (Bishop 2006). Suppose a ML model is tasked with learning a mapping from a set of input variables $\boldsymbol{x} \equiv \{x_1, \dots, x_n\}$ to a set of output variables $\boldsymbol{y} \equiv \{y_1, \dots, y_m\}$. In practice, this is undertaken by training the ML model on a finite number of examples, *i.e.*, a dataset $\left\{\left(\boldsymbol{x}^{(i)}, \boldsymbol{y}^{(i)}\right)\right\}_{i=1}^N$ where $i = 1, \dots, N$ and $N$ is the total number of observations in the dataset. Let observation 1 have input variables $\boldsymbol{x}^{(1)}$ and output variables $\boldsymbol{y}^{(1)}$, and observation 2 have input variables $\boldsymbol{x}^{(2)}$ and output variables $\boldsymbol{y}^{(2)}$. Suppose the following one-to-many relation exists: $\boldsymbol{x}^{(1)} = \boldsymbol{x}^{(2)} \equiv \boldsymbol{x}^{(a)}$ and $\boldsymbol{y}^{(1)} \neq \boldsymbol{y}^{(2)}$ as shown in **figure 1**. The ML model would consequently have to learn a non-unique relation between $\boldsymbol{x}$ and $\boldsymbol{y}$ for $\boldsymbol{x} = \boldsymbol{x}^{(a)}$. This causes difficulty in training because the loss function would use $\boldsymbol{y}^{(1)}$ and $\boldsymbol{y}^{(2)}$ alternatingly as the target value to calculate error in the predicted value of $\boldsymbol{y}$ for $\boldsymbol{x} = \boldsymbol{x}^{(a)}$. Therefore, the ML model would never reach a converged state – resulting in a poor fitting, and it would always give erroneous predictions for $\boldsymbol{y}$ at $\boldsymbol{x} = \boldsymbol{x}^{(a)}$ (Bishop 1994; Liu *et al.* 2018).

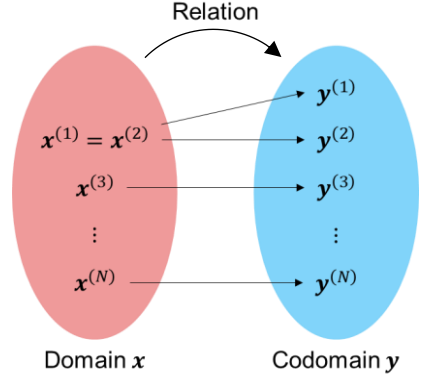

**Figure 1** Diagram showing the mapping from inputs domain $\boldsymbol{x}$ to outputs codomain $\boldsymbol{y}$. A non-unique one-to-many relation is shown between $\boldsymbol{x}$ and $\boldsymbol{y}$ at $\boldsymbol{x} = \boldsymbol{x}^{(1)}\left(= \boldsymbol{x}^{(2)}\right)$, while a unique one-to-one relation is shown between $\boldsymbol{x}$ and $\boldsymbol{y}$ at $\boldsymbol{x} = \boldsymbol{x}^{(3)}$ and $\boldsymbol{x} = \boldsymbol{x}^{(N)}$.

### 2.3    Non-Unique Mapping from 1D Channel Flow Data

It has been reported in the literature that NUM can occur in TBML models when trained on cases where shear velocity gradient $\partial \bar{u}_1 / \partial x_2$ is the only non-negligible velocity gradient, such as in fully-developed channel flow (Liu *et al.* 2021). In this situation, the expressions for the $b_{ij}$ components given by the GEVH reduce to:

$$\begin{aligned} b_{11} &= f_1 \\ b_{22} &= f_2 \\ b_{33} &= -(f_1 + f_2) \\ b_{12} &= f_3 \end{aligned} \tag{2.5}$$

where $f_1, f_2$, and $f_3$ are univariate functions of nondimensional shear velocity gradient, $\alpha \left(= (k/\varepsilon)(\partial \bar{u}_1 / \partial x_2)\right)$, due to the $g_n$ coefficients and basis tensors becoming univariate functions of $\alpha$. Liu *et al.* (2021) showed that for any value of $\alpha$ present in DNS data for 1D channel flow, there are multiple possible values of $b_{12}$. This also occurs in the normal $b_{ij}$ components as shown by Cai *et al.* (2022). Given that $\alpha$ in channel flow simulated using RANS and scale-resolved methods have a similar profile, a TBML model trained on channel flows would experience NUM. This is because the model by virtue of Eq. (2.5) would be tasked with mapping $\alpha$ from RANS to multiple true values from a scale-resolved method for each $b_{ij}$ component (Wilcox 2006). To illustrate this, **figure 2(a)** shows a plot of $\alpha$ given by a two-equation RANS model against $b_{12}$ given by well-resolved LES for a channel flow at $Re_\tau = 945$ (based on friction velocity and channel half-height). It is clear that a NUM exists between $\alpha$ and $b_{12}$ in **figure 2(a)**. The annotations show that distance from the wall has a many-to-one relation with $b_{12}$. Therefore, some studies have proposed sensitising $b_{12}$ to a supplementary input variable (SIV) that monotonically changes with wall distance to address this issue. One such variable is the turbulent Reynolds number $Re_t \left(= k^2 / \nu \varepsilon\right)$, which was included as an SIV by Liu *et al.* (2021) and Jiang *et al.*

(2021). By plotting **figure 2(a)** with $Re_t$ on a third axis as shown in **figure 2(b)**, it is seen that a unique mapping exists between input variables $\{\alpha, Re_t\}$ and output variable $b_{12}$.

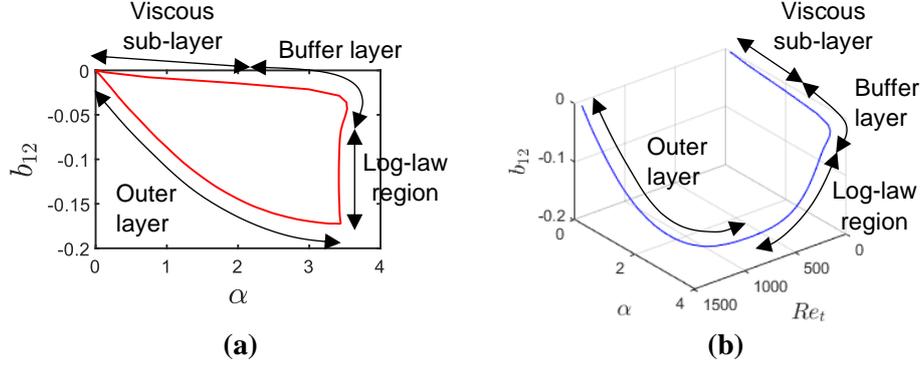

**(a)**          **(b)**

**Figure 2** Fully-developed turbulent channel flow at $Re_\tau = 945$ (based on friction velocity and channel height) with $\alpha$ and $Re_t$ given by a two-equation RANS approach and $b_{12}$ given by well-resolved LES: **(a)** a non-unique mapping between $\alpha$ and $b_{12}$ is shown, and **(b)** a unique mapping between input variables $\{\alpha, Re_t\}$ and output variable $b_{12}$ is shown.

## 2.4    2D General Effective Viscosity Hypothesis Equations

While NUM due to 1D channel flow data in tensor-basis machine learning (TBML) models has been well studied, it has not yet been properly investigated with 2D flows, even though data from such cases has been mostly used for training TBML models. In a related study, Sotgiu *et al.* (2019) briefly discussed about NUM caused by data from a 2D corrugated channel with dimensionless wall-normal Reynolds stress $v^2/k$ as one of two inputs. However, as $v^2/k$ is seldom used in TBML models, their discussion is not generally applicable. Therefore, the present work aims to give a detailed investigation on NUM with the standard invariant inputs, which are used in most TBML models proposed in the literature. A natural starting point is to examine the expressions and variables in the 2D GEVH. The component-wise form of the $b_{ij}$ expression given by the GEVH in Eq. (2.2) can be simplified as:

$$b_{11} = g_1 S_{11} - 2g_2 S_{12} R_{12} + \frac{1}{3} g_3 (S_{11}^2 + S_{12}^2)$$
$$b_{22} = -g_1 S_{11} + 2g_2 S_{12} R_{12} + \frac{1}{3} g_3 (S_{11}^2 + S_{12}^2)$$
$$b_{33} = -\frac{2}{3} g_3 (S_{11}^2 + S_{12}^2) \tag{2.6}$$
$$b_{12} = g_1 S_{12} + 2g_2 S_{11} R_{12}$$

and the components of nondimensional mean strain and rotation rate are calculated as:

$$S_{11} = -S_{22} = \frac{k}{\varepsilon} \frac{\partial \bar{u}_1}{\partial x_1}, \qquad S_{12} = S_{21} = \frac{k}{2\varepsilon} \left( \frac{\partial \bar{u}_1}{\partial x_2} + \frac{\partial \bar{u}_2}{\partial x_1} \right),$$
$$R_{12} = -R_{21} = \frac{k}{2\varepsilon} \left( \frac{\partial \bar{u}_1}{\partial x_2} - \frac{\partial \bar{u}_2}{\partial x_1} \right)$$

$g_1$, $g_2$ and $g_3$ are functions of $tr(\boldsymbol{S}^2)$ and $tr(\boldsymbol{R}^2)$ as stated in Eq. (2.4):

$$tr(\boldsymbol{S}^2) = 2(S_{11}^2 + S_{12}^2)$$
$$tr(\boldsymbol{R}^2) = -2R_{12}^2 \tag{2.7}$$

The combination of $tr(\boldsymbol{S}^2)$ and $tr(\boldsymbol{R}^2)$ as a set of input variables will be denoted as $\boldsymbol{x}_{tr}$, such that $\boldsymbol{x}_{tr} \equiv \{tr(\boldsymbol{S}^2), tr(\boldsymbol{R}^2)\}$. As the 2D version of the GEVH is examined in the rest of this paper, $g_n$ will collectively represent $g_1$, $g_2$, and $g_3$ hereafter, such that $n = (1, 2, 3)$ unless stated otherwise. Given that the full GEVH with ten basis tensors reduces to this version when it is evaluated on 2D flows, the results and conclusions of this study are also applicable to TBML models based on the full GEVH. Eq. (2.6) will only be studied in the context of TBML models, so the $b_{ij}$ components on the left-hand side will be considered as those of true $b_{ij}$, with $\boldsymbol{S}$ and $\boldsymbol{R}$ components from RANS on the right-hand side.

## 3    Methodology

### 3.1    Inputs and Outputs for Non-Unique Mapping Analysis

Appropriate input and output variables must be chosen to investigate whether NUM can exist in the 2D GEVH and subsequently, in TBML models trained on 2D flow data. For NUM analysis of 1D channel flow data, $\alpha$ and true $b_{ij}$ have usually been chosen as the input and output variables, respectively, because the true $b_{ij}$ components become univariate functions of $\alpha$ as shown in Eq. (2.5) (Liu *et al.* 2021, Cai *et al.* 2022). In the GEVH, the inputs are $\boldsymbol{x}_{tr}$ so these should be chosen as the NUM analysis inputs. For the outputs, the $b_{ij}$ components are found to be functions of $S_{11}$, $S_{12}$, $R_{12}$, $g_1$, $g_2$, and $g_3$ as shown in Eq. (2.6). While it seems reasonable to choose true $b_{ij}$ again as the outputs for NUM analysis, doing so ignores the specific operations on $g_n$ in the tensor-basis. This can lead to false diagnoses of NUM in a TBML model.

Consider the following scenario where two arbitrary locations denoted L1 and L2 in a 2D flow case have the same $R_{12}$ values but different $S_{11}$ and $S_{12}$ values that satisfy the condition: $S_{11,L1}^2 + S_{12,L1}^2 = S_{11,L2}^2 + S_{12,L2}^2$, where $S_{11,L1} \neq S_{11,L2}$, $S_{12,L1} \neq S_{12,L2}$, and subscripts L1 and L2 denote the location each term belongs to. This leads to both locations having the same $\boldsymbol{x}_{tr}$ input values. If L1 and L2 also have the same true $b_{ij}$ values, then a one-to-one relation would exist between $\boldsymbol{x}_{tr}$ and true $b_{ij}$. However, L1 and L2 require different true $g_n$ values to be predicted to give the same true $b_{ij}$ values, as L1 and L2 would apply different $S_{11}$ and $S_{12}$ values to Eq. (2.6). Hence, a one-to-many relation would exist between $\boldsymbol{x}_{tr}$ and true $g_n$. This relation would cause NUM in TBML models because while the final output in TBML models is $b_{ij}$, their learnable mapping actually exists between $\boldsymbol{x}_{tr}$ and $g_n$. Therefore, this NUM is undetectable if true $b_{ij}$ is chosen as the NUM analysis output and would give a false negative result. Oppositely, it is possible that a one-to-one relation between $\boldsymbol{x}_{tr}$ and true $g_n$ may exist if L1 and L2 have different true $b_{ij}$ values instead, thus returning a false positive result.

NUM in TBML models is guaranteed to be absent if one-to-many relation between $\mathbf{x}_{tr}$ and true $g_n$ do not exist – allowing both $g_n$ and $b_{ij}$ to be predicted accurately. The task becomes finding the true $g_n$ values. McConkey *et al.* (2022) showed that true $g_n$ can be approximated from a least-squares solution between both sides of Eq. (2.6). Mandler and Weigand (2022) obtained true $g_n$ by performing field inversion involving successive tensor projections. However, both methods do not express true $g_n$ explicitly with closed-form equations, which would allow the causes of one-to-many relation between $\mathbf{x}_{tr}$ and true $g_n$ to be more easily understood and addressed. Jongen and Gatski (1998) showed that true $g_n$ can be explicitly determined with the following expressions which are always valid for 2D flows:

$$g_1 = \left(\frac{S_{11}}{2(S_{11}^2 + S_{12}^2)}\right)(b_{11} - b_{22}) + \left(\frac{S_{12}}{S_{11}^2 + S_{12}^2}\right)b_{12} \tag{3.1a}$$

$$g_2 = \frac{2S_{11}b_{12} + S_{12}(b_{22} - b_{11})}{4R_{12}(S_{11}^2 + S_{12}^2)} \tag{3.1b}$$

$$g_3 = \frac{3(b_{11} + b_{22})}{2(S_{11}^2 + S_{12}^2)} \tag{3.1c}$$

Although they derived Eq. (3.1) by finding the Gram matrix for calculating true $g_n$, these expressions can be obtained by treating Eq. (2.6) as a set of simultaneous equations and solving for $g_1$, $g_2$, and $g_3$.

As a side note: while these true coefficients can be used as the targets in TBML models, the calculation of them can also be used to tune well-established RANS models. The two-equation RANS models such as the $k - \varepsilon$ model are known to have accuracy limitations due to $C_\mu$ being set as a constant in calculating Reynolds stress. Coefficient $g_1$ is directly proportional to $C_\mu$ and therefore calculating true $g_1$ may allow target values of $C_\mu$ to be calculated for tuning the models. Furthermore, calculating true $g_2$ and $g_3$ may allow developers working on these classical models to add accurate terms to Boussinesq hypothesis to increase the accuracy of their model.

### 3.2    True $g_n$ as the Non-Unique Mapping Analysis Output

Eq. (3.1a-c) gives the true $g_n$ values for any 2D flow case, and a TBML model achieves the upper performance limit if it can predict these from inputs $\mathbf{x}_{tr}$. Hence, true $g_n$ should be the $g_n$ targets for a TBML model to predict and therefore chosen as the NUM analysis output. There is a caveat however if the *a posteriori* process is to be performed, in which test predictions from the TBML model are injected into the RANS equations to give improved mean flow field results. Wu *et al.* (2019) proposed that the linear term should be treated implicitly and give a non-negative eddy viscosity to ensure numerical stability, *i.e.*, $\nu_t \geq 0$, which can only be achieved if $g_1 \leq 0$ (Durbin & Pettersson-Reif 2011; McConkey *et al.* 2022). The TBML model should be trained to predict $g_1 \leq 0$ to enforce this constraint, which requires $g_1$ targets $\leq 0$ in the training data. As true $g_n$ calculated with Eq. (3.1) are not constrained to obey this inequality, they should not be the training targets in this situation where true $g_1 > 0$. Therefore, at these locations, different $g_n$ targets would need to be considered that (i) obey $g_1$ targets $\leq 0$, (ii) return true $b_{ij}$ when substituted into the GEVH in Eq. (2.6), and (iii) ideally are given by closed-form expressions so that any causes of NUM can be traced back to the flow physics.

Nonetheless, true $g_n$ should always be the $g_n$ targets where true $g_1 \leq 0$. Therefore, if true $g_1 \leq 0$ occurs in most of the domain of a training case, choosing true $g_n$ as the NUM analysis output would be applicable to most of the data, and can still give many insights on the mapping between the inputs $\boldsymbol{x}_{tr}$ and true $g_n$. With data from such flow cases, choosing true $g_n$ as the NUM analysis output would allow the main aim of this study to be achieved, which was to identify whether NUM occurs due to 2D flow data. With this reason in mind, true $g_n$ was chosen as the NUM analysis output, and data from flow cases that mostly contained values of true $g_1 \leq 0$ which are introduced in the next section were used.

### 3.3    Flow Cases

The NUM analysis was performed on data from two different flow cases: (i) a flow over periodic hills characterised by the interaction of the fluid with repetitive hill-like structures, influencing boundary layer dynamics, flow separation, and turbulence generation, and (ii) a plane impinging jet involving various flow patterns such as stagnation, boundary layer effects, and recirculation zones. These cases were chosen to demonstrate the developed methodology on different 2D flow physics. Velocity contours of these cases with annotations are shown in **figure 3(a)** and **(b)**.

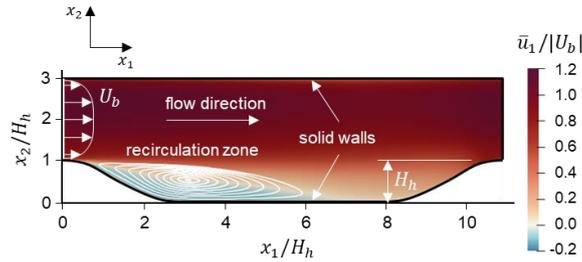

**(a)** Flow over periodic hills at $Re = 5600$, based on bulk inlet velocity of 0.028 m/s and hill crest height of 1 m.

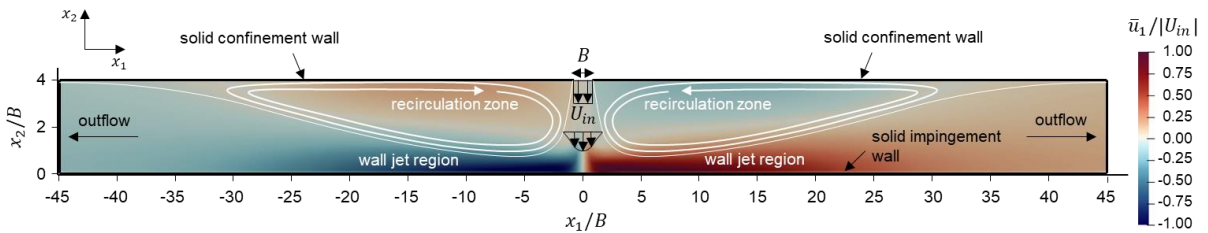

**(b)** Plane impinging jet flow at $Re = 2 \times 10^4$ based on bulk inlet velocity of 3 m/s and slot width of 0.1 m.

**Figure 3** Schematic velocity contour plots of **(a)** flow over the periodic hills case, and **(b)** the impinging jet case.

The 2D flow over periodic hills case is a wall-bounded flow which separates at the hilltop, creating a large recirculation zone with flow reattachment downstream of the hill. The Reynolds number is 5600 based on a bulk inlet velocity $U_b$ of 0.028 m/s and a hill crest height $H_h$ of 1 m. It has a steepness factor, $\beta$ of 1.5 which gives a domain length $L_x$ of $10.9H_h$ as shown in **figure 3(a)**. Further information regarding the geometry and boundary conditions can be found in Xiao *et al.* (2020). Data was extracted from McConkey *et al.* (2021), which includes inputs $x_{tr}$ modelled with RANS shear stress transport (SST), and true $b_{ij}$ from a DNS performed by Xiao *et al.* (2020). This case was chosen for NUM analysis because periodic hill cases are commonly used to train TBML models in the literature (Kaandorp & Dwight 2020; McConkey *et al.* 2022; Tang *et al.* 2023).

To assess NUM on a very different canonical flow case without flow separation, data from an impingement flow was also used. As $x_{tr}$ and true $b_{ij}$ data of an impingement flow was not readily available, a RANS and scale-resolved simulation of one was run in this study. Jaramillo *et al.* (2012) and Shukla and Dewan (2019) simulated a particular plane impinging jet case using RANS and scale-resolved methods. This case begins with uniform velocity flow entering through a slot in the top wall of the domain, which forms a core jet as it falls towards the bottom. This leads to an impingement zone due to stagnation at the bottom wall. The flow then moves towards the sides to form wall jets, counter-rotating recirculation zones, and leaves the domain at the two side outlets. The Reynolds number is $2 \times 10^4$ (based on bulk inlet velocity $U_{in}$ and slot width $B$), and the spacing between the inlet and impingement wall is $4B$ as shown in **figure 3(b)**. In this work, the case was modelled in OpenFOAM version 2006 with RANS SST and well-resolved LES to obtain inputs $x_{tr}$, and spanwise- and time-averaged true $b_{ij}$, respectively (Jasak *et al.* 2007; Menter 1993; Kim & Menon 1995). Validation of the results was performed by comparing the recirculation zone dimensions with the two studies, and good agreement was obtained. For both the periodic hill and impinging jet cases, true $b_{ij}$ was interpolated onto their respective RANS grids. The $S$ and $R$ components from RANS and interpolated true $b_{ij}$ were then used to calculate true $g_n$ at each RANS cell centre with Eq. (3.1).

### 3.4 Clustering Process

Quantifying how NUM varies in the input space allows the worst affected regions to be determined, and more effective methods to reduce or eliminate NUM can be developed by targeting these regions. The challenge that 2D flow data poses is that $x_{tr}$ vs. true $g_n$ ($n \in \{1, 2, 3\}$) exists in three dimensions. One suggestion would be to extend the method used for visualising NUM in 1D channel flow data as shown in **figure 2** by creating a surface plot with $x_{tr}$ on the $x$ and $y$ axis and true $g_n$ ($n \in \{1, 2, 3\}$) on the $z$-axis. Overlapping surfaces in regions of the $x_{tr}$ space would then indicate a one-to-many relation between $x_{tr}$ and the plotted true $g_n$ coefficient (Shizawa 1994). Although the amount of input space that contains overlapping surfaces can be easily quantified when inputs $x_{tr}$ are used, this becomes challenging with three or more inputs, and comparison becomes unclear, *e.g.*, comparing overlapping surfaces to overlapping volumes. Secondly, many overlapping surfaces can lead to cluttered data visualisations and difficulties in clearly displaying regions of the input space that would be worst affected by NUM.

The proposed NUM quantification method is instead based on inputs vs. output data in the form of scatter plots with inputs $\boldsymbol{x}_{tr}$ plotted on the $x$ and $y$ axes, and outputs true $g_n$ ($n \in \{1, 2, 3\}$) given in colour. For each scatter plot, the method begins by grouping points with similar true $g_n$ ($n \in \{1, 2, 3\}$) values into a pre-specified number of clusters. Each data point can then be compared with other points that have very similar $\boldsymbol{x}_{tr}$ values to assess whether any of them belong to a different cluster. If so, this indicates that the two points have very similar $\boldsymbol{x}_{tr}$ but very different true $g_n$ ($n \in \{1, 2, 3\}$) values, signifying that NUM would exist at those $\boldsymbol{x}_{tr}$ values. This approach is based on finding regions of the input space where groups of points with similar true $g_n$ ($n \in \{1, 2, 3\}$) values overlap (Huang *et al.* 2013). Complete linkage agglomerative clustering in the Scikit-Learn library was used to group the points (Pedregosa *et al.* 2011). This algorithm begins by considering each point as a cluster of its own, then clusters with closest similarity are combined into larger ones until a pre-specified number of clusters remain (Gan *et al.* 2007). More information regarding the clustering process can be found in **Appendix A**.

To find the ideal number of clusters, a cluster validity study was performed whereby the clustering process was run with the pre-specified number of clusters set to $2, 3, 4, \ldots, 10$. The ideal number was chosen as (i) greater than two, as it was found that two clusters still contained a large range of true $g_n$ ($n \in \{1, 2, 3\}$) values, and (ii) the value that gave the highest average silhouette score over all true $g_n$ coefficients (Aggarwal & Reddy 2014). The average silhouette score is a commonly used metric of cluster quality that represents how separated the clusters are from each other and how compact they are. This is achieved by considering both the intra- and inter-cluster distances between points. Readers are referred to Rousseeuw (1987) for further information. **Figure 4** shows that the ideal number of clusters was 3 for both cases.

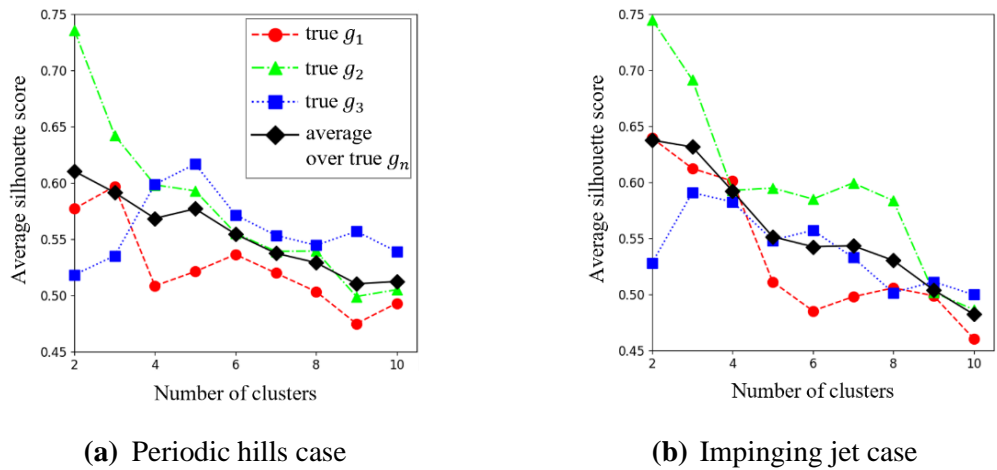

**(a)** Periodic hills case        **(b)** Impinging jet case

**Figure 4** Line plots showing number of clusters vs. silhouette score for true $g_n$ ($n \in \{1, 2, 3\}$) and their average for the **(a)** periodic hill case, and **(b)** impinging jet case.

### 3.5 Non-Unique Mapping Quantification Process

For any chosen point in a cluster denoted hereafter as the "focal point", two approaches named the nearest neighbours method (NNM), and proximity box method (PBM) were undertaken to identify points that have very similar $x_{tr}$ values to it, which are hereafter referred to as "close points". These approaches are illustrated in **figure 5(a)** and **(b)**, respectively. In NNM, the nearest $n$ points are taken as the close points, where $n$ is a pre-specified number (Papadopoulos & Manolopoulos [2005]). In this study, $n$ was set to the number of clusters, which ensured each cluster had a chance to allocate at least one of their points as a nearest neighbour. The $x_{tr}$ distance between any two points $p$ and $q$ was determined with Euclidean distance, $d$:

$$d = \sqrt{\left(tr(\boldsymbol{S}^2)_p - tr(\boldsymbol{S}^2)_q\right)^2 + \left(tr(\boldsymbol{R}^2)_p - tr(\boldsymbol{R}^2)_q\right)^2} \tag{3.2}$$

where the subscripts indicate which point the trace is calculated from. In PBM, a box with length $l$ and height $h$ is specified (de Berg *et al.* [2000]). The box is then centrally superimposed on the focal point, and the other points inside the box are taken as the close points. NNM and PBM are similarity query methods, which are well-established in data similarity searching (Zezula *et al.* [2006]).

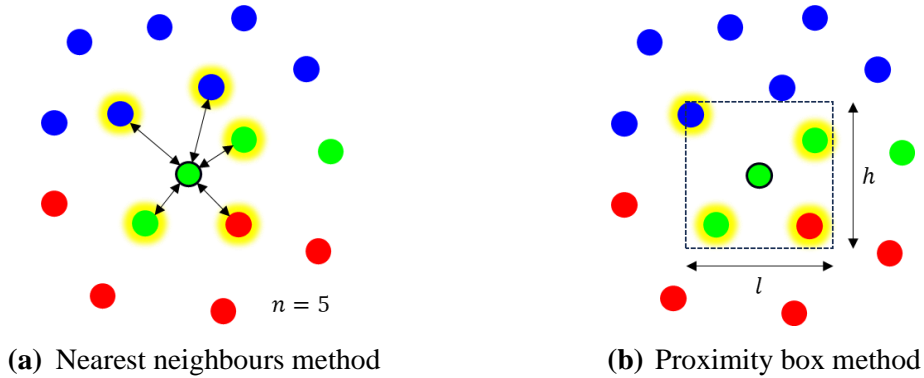

**(a)** Nearest neighbours method        **(b)** Proximity box method

**Figure 5** Close points identification methods using **(a)** nearest neighbours, and **(b)** a proximity box. The focal point has a black border and close points are highlighted in yellow.

If the focal point and one of its close points belong to different clusters, then a one-to-many relation is considered to exist between their $x_{tr}$ values and the true $g_n$ coefficient being investigated. This will be referred to as a "conflicting instance" (CI), which is proposed as a metric to quantify NUM in this work. The distribution of CIs in the input space was investigated by partitioning the $x_{tr}$ space into a grid, sequentially setting each scatter point in each grid cell as the focal point and calculating their number of CIs, then summing them to give the total number of CIs in each grid cell, which is denoted as $n_{CI}$. This process is outlined in **Algorithm 1** and was separately run for true $g_n$, *i.e.*, $\forall n \in \{1, 2, 3\}$.

---
**Algorithm 1:** Find distribution of CIs as a metric of NUM in the input space

---

1    Partition the input space into a grid
2    **for** each grid cell $c = 1, 2, …, N_c$ **do**
3      Initialize new CI counter $n_{CI} = 0$ for grid cell $c$
4      **for** each point in grid cell $p_c = 1, 2, …, N_p^c$ **do**
5        Run NNM or PBM to find its close points $p_c' = 1, 2, …, N_p^{c'}$
6        **for** each close point $p_c' = 1, 2, …, N_p^{c'}$ **do**
7          **if** focal point $p_c$ and close point $p_c'$ belong to different clusters **then**
8          $n_{CI} ← n_{CI} + 1$
9          **end if**
10        **end for**
11      **end for**
12    **end for**

---

This method can be extended to include more than two input variables by performing the clustering with more dimensions and increasing the dimensions of Eq. (3.2) if using NNM, or applying a higher dimensional proximity space (*e.g.*, cuboid for three inputs) if using PBM. $n_{CI}$ can thereby be easily calculated and compared to assess how NUM changes when using a greater number of, or different input variables.

As an introductory demonstration, the method was applied to 1D channel flow containing input variable $\alpha$ and the results are detailed in **Appendix B**. While the present approach may be extended to 3D flows, it is not possible for the true output coefficients to be calculated analytically as was undertaken in this study with Eq. (3.1). Instead, they must be calculated numerically – *e.g.*, by least squares or successive tensor projections (McConkey *et al.* 2022; Mandler & Weigand 2022). Furthermore, a minimum of five invariant inputs are required. This method is also not limited to TBML models; it can also be applied to other data-driven turbulence models where the inputs and target outputs of the training cases are well-defined. With the two typical TBML model inputs in $\boldsymbol{x}_{tr}$, this method enables NUM caused by 2D flow data to be clearly visualised and quantified as demonstrated in the following results.

## 4    Periodic Hills Case Results

### 4.1    Non-Unique Mapping Quantification

The NUM quantification results for the periodic hills case are shown in **figure 6**. It is observed in **figure 6(a)-(c)** that the data points are grouped according to their true $g_n$ ($n \in \{1, 2, 3\}$) values, and the groups overlap in some parts of the $\boldsymbol{x}_{tr}$ space. This can be easily visualised in the low strain and rotation (LSR) subset of $\boldsymbol{x}_{tr}$, which is shown in the zoomed-in views. For example in **figure 6(a)**, true $g_1$ at $tr(\boldsymbol{S}^2) \approx 2$ and $tr(\boldsymbol{R}^2) \approx -1$ is approximately both greater than $0.02$ and less than $-0.1$. This is also observed with true $g_2$ and true $g_3$ for the same input values, thereby qualitatively showing that NUM can exist between $\boldsymbol{x}_{tr}$ and true $g_n$ in 2D flow data. Another part of the $\boldsymbol{x}_{tr}$ space where there are overlapping groups (and thus NUM can exist) is near and along the $tr(\boldsymbol{S}^2) = -tr(\boldsymbol{R}^2)$ diagonal. These points were taken from locations that have approximately pure or homogeneous shear flow (Taghizadeh *et al.* 2021; Mishra & Girimaji 2013).

A comparison between **figure 6(a)-(c)** and **figure 6(d)-(f)** shows that the scatter points have been reasonably clustered according to their true $g_n$ values. These clusters enabled the $n_{CI}$ heat maps calculated with NNM shown in **figure 6(g)-(i)** to be made, which support the scatter plot observations. In particular, the heat maps show that there is high $n_{CI}$ located in the LSR subset where $tr(\boldsymbol{S}^2) < 2$ and $tr(\boldsymbol{R}^2) > -2$ and near the pure shear line, amounting to 77%, 76% and 74% of total $n_{CI}$ in the true $g_1$, $g_2$, and $g_3$ heat maps, respectively, found in the heat map grid cells that the pure shear line passes through. While it is difficult to identify NUM in the rest of the $\boldsymbol{x}_{tr}$ space using the scatter plots in **figure 6(a)-(c)**, the heat maps show that CIs exist there but in significantly lower quantities. **Figure 17(a)-(c)** in **Appendix C** shows that the distribution trends of $n_{CI}$ given by **Algorithm 1** with PBM are very similar to those of **figure 6(g)-(i)**, and therefore the results of both approaches support each other.

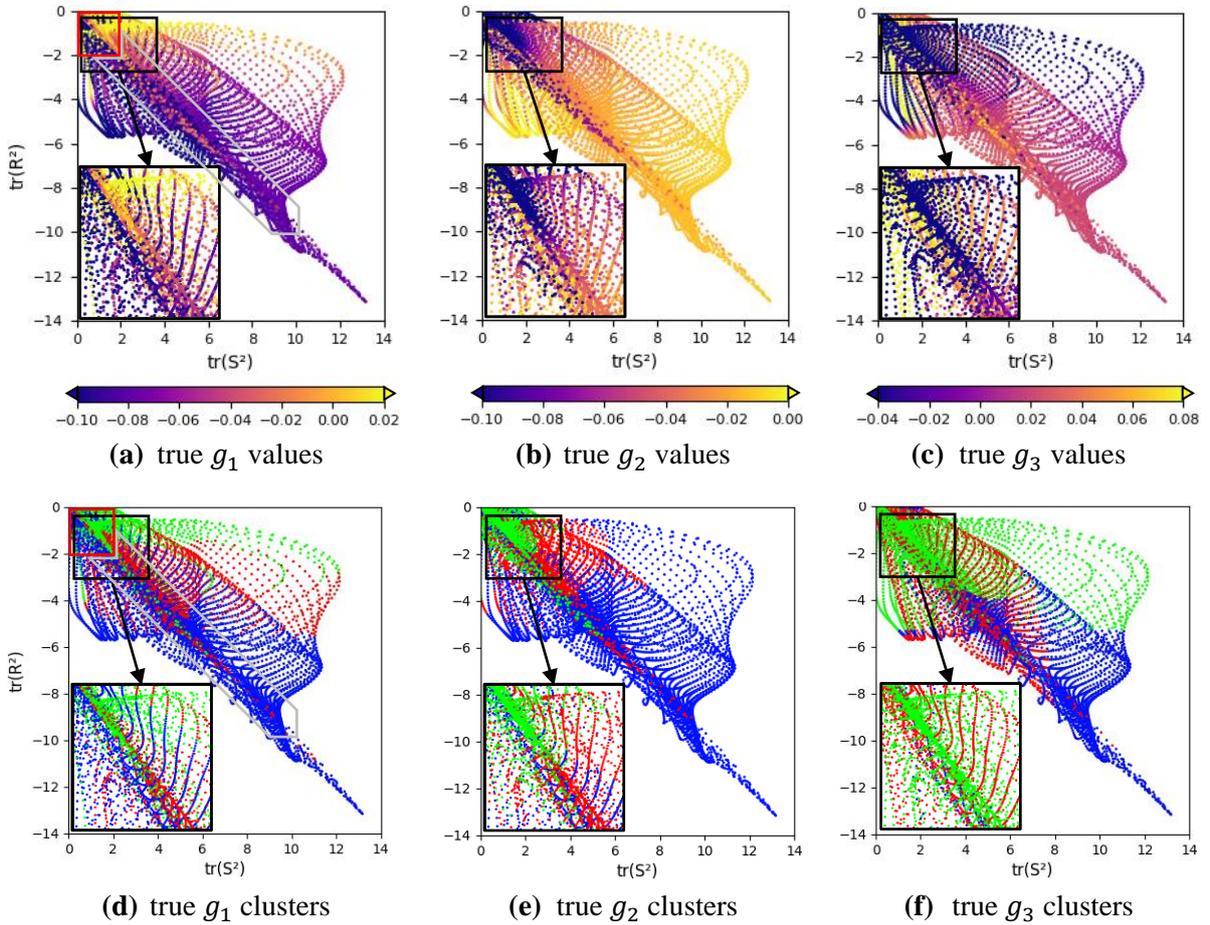

**(a)** true $g_1$ values  **(b)** true $g_2$ values  **(c)** true $g_3$ values

**(d)** true $g_1$ clusters  **(e)** true $g_2$ clusters  **(f)** true $g_3$ clusters

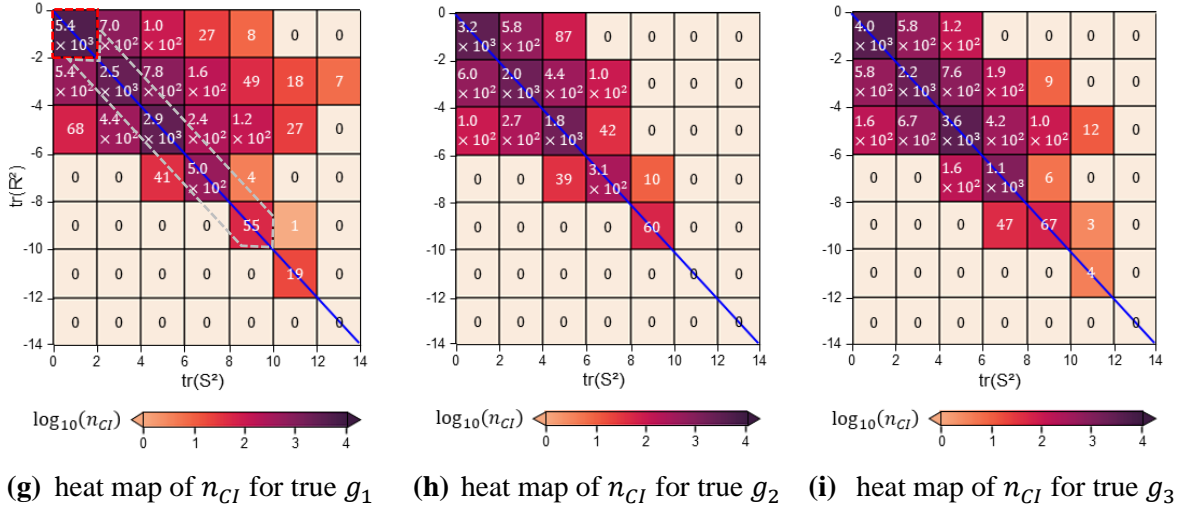

**(g)** heat map of $n_{CI}$ for true $g_1$  **(h)** heat map of $n_{CI}$ for true $g_2$  **(i)** heat map of $n_{CI}$ for true $g_3$

**Figure 6** NUM results for the flow over periodic hills data. Subplots **(a)-(c)** show scatter plots of inputs $x_{tr}$ vs. outputs true $g_n$, **(d)-(f)** show clusters represented by colours: red, blue, and green, and **(g)-(i)** show heat maps of $n_{CI}$ given by NNM ($n = 3$), with the blue line indicating pure shear flow. The red boxes in subplots **(a)**, **(d)**, and **(g)** mark the low strain and rotation subset, and the grey boxes mark the near pure shear subset described in the following two sections.

## 4.2 Low Strain and Rotation Regions

To relate the findings from **figure 6** to the physical domain, contours of the NUM analysis inputs and outputs are shown in **figure 7(a)-(b)** and **figure 7(c)-(e)**, respectively. The true $g_n$ contour distributions in **figure 7(c)-(e)** are found to compare well with optimal $g_n$ computed using field inversion in Mandler and Weigand (2023), thereby validating the approach of using Eq. (3.1) to calculate true $g_n$.

The domain regions in **figure 7** with red hatching represent where scatter points located in the LSR subset of $tr(S^2) < 2$ and $tr(R^2) > -2$ in **figure 6** were taken from. Two regions that satisfy these constraints can be identified: the larger region in the freestream and the smaller region by the lower wall – both spanning the whole length of the domain. The constraints can be rewritten in terms of velocity gradients by substituting Eq. (2.7) into them:

$$\left(\frac{\partial \bar{u}_1}{\partial x_1}\right)^2 + \frac{1}{4}\left(\frac{\partial \bar{u}_1}{\partial x_2} + \frac{\partial \bar{u}_2}{\partial x_1}\right)^2 < \frac{\varepsilon^2}{k^2}$$

$$\frac{1}{4}\left(\frac{\partial \bar{u}_1}{\partial x_2} - \frac{\partial \bar{u}_2}{\partial x_1}\right)^2 < \frac{\varepsilon^2}{k^2}$$

(4.1)

The presence of $\partial \bar{u}_1 / \partial x_2$ in both expressions of Eq. (4.1) and its large range compared to other velocity gradients suggests that it is important in determining whether a data point belongs to red hatching or not in this case. It is found that the red hatched regions in **figure 7** contain inflection points of $\partial \bar{u}_1 / \partial x_2$ (Xiao *et al.* 2020). Therefore, the LSR flow is found to have only been extracted from regions with very low values of $\partial \bar{u}_1 / \partial x_2$ – approximately one order of magnitude lower than the rest of the domain.

**Figure 7(c)** shows discontinuities exist in true $g_1$ as marked by label (1). These are caused by terms in the true $g_1$ expression in Eq. (3.1a) tending towards infinity or negative infinity as $S_{11}$ and $S_{12}$ approach zero. Consequently, low $S_{11}^2 + S_{12}^2$ values (on the order of $10^{-3}$ to $10^{-1}$ in this case) can cause discontinuities, such that true $g_1$ may change from high magnitude negative values to high positive values across them. Discontinuities in true $g_2$ and $g_3$ can also be observed as marked by label (2) in **figure 7(d)** and **figure 7(e)**, respectively. These are also caused by terms tending towards infinity and negative infinity as $S_{11}$ and $S_{12}$ approach zero. Proof of these limits are given in **Appendix D**. Negligible change in $tr(\boldsymbol{S}^2)$ and $tr(\boldsymbol{R}^2)$ can be observed in **figure 7(a)-(b)** where the discontinuities occur. This shows that one-to-many relations exist between $\boldsymbol{x}_{tr}$ and true $g_n$ at these $\boldsymbol{x}_{tr}$ values. The clustering was able to capture the large differences in true $g_n$ across these discontinuities, resulting in high $n_{CI}$ in the LSR subset at $tr(\boldsymbol{S}^2) < 2$ and $tr(\boldsymbol{R}^2) > -2$ as shown in **figure 6(g)-(i)**.

However, it is found that NUM from the discontinuities has a negligible effect on the accuracy of $b_{ij}$ given by TBML models. For example, consider a TBML model that is able to perfectly predict the true $g_n$ fields shown in **figure 7(c)-(e)**. The model would then typically multiply its $g_n$ predictions (which are true $g_n$ in this scenario) with $\boldsymbol{S}$ and $\boldsymbol{R}$ tensor products in the GEVH to give $b_{ij}$ predictions as shown in Eq. (2.6). It is found that any discontinuities in $g_n$ vanish upon substitution in Eq. (2.6), and negligible differences in $b_{ij}$ are obtained whether $g_n$ have high negative or positive magnitude predictions in regions containing low $S_{11}$ and $S_{12}$. Further explanations are given in **Appendix E**. This indicates that as long as the loss function in the TBML model is calculated using $b_{ij}$ error, NUM caused by true $g_n$ discontinuities in these regions can be ignored.

### 4.3    Near Pure Shear Regions

The white hatched regions in **figure 7** represent where scatter points in **figure 6** that were not included in the red hatching and satisfy the following constraints were taken from:

$$-1 < tr(\boldsymbol{S}^2) + tr(\boldsymbol{R}^2) < 1$$
$$tr(\boldsymbol{S}^2) < 10 \tag{4.2}$$
$$tr(\boldsymbol{R}^2) > -10$$

The first constraint captures points that are in close proximity to the pure shear line, while the second and third constraint excludes points in a region of the $\boldsymbol{x}_{tr}$ space that has nil or very low $n_{CI}$, as shown in **figure 6(g)-(i)**. Substituting Eq. (2.7) into the first constraint shows that these points satisfy:

$$-\frac{\varepsilon^2}{2k^2} < \left(\frac{\partial \bar{u}_1}{\partial x_1}\right)^2 + \frac{\partial \bar{u}_1}{\partial x_2}\frac{\partial \bar{u}_2}{\partial x_1} < \frac{\varepsilon^2}{2k^2} \tag{4.3}$$

and in the same way, the criterion for pure shear, $tr(\boldsymbol{S}^2) = -tr(\boldsymbol{R}^2)$ can be found as:

$$\left(\frac{\partial \bar{u}_1}{\partial x_1}\right)^2 = -\frac{\partial \bar{u}_1}{\partial x_2}\frac{\partial \bar{u}_2}{\partial x_1} \qquad (4.4)$$

There are two main regions of white hatching in **figure 7**: (i) the region at the centre of the domain, which includes freestream, recirculation, and reattachment flow, and (ii) the near-wall region by the upper wall. This shows that different flow physics can satisfy Eq. (4.3) to be considered flow that is almost pure shear. These different flow physics are found to cause NUM between $\boldsymbol{x}_{tr}$ and true $g_n$ in various ways. For example, some regions have almost constant values of $tr(\boldsymbol{S}^2)$ and $tr(\boldsymbol{R}^2)$ but contain a large range of true $g_n$, such as near the separation point labelled (3) in **figure 7**. A special case of this occurs where there is pure shear flow, such that $tr(\boldsymbol{S}^2) = -tr(\boldsymbol{R}^2)$. As one input is equal to the negative of the other, the number of inputs effectively reduces to one at these locations. **Figure 7** shows that some pure shear flow regions have almost constant values of $tr(\boldsymbol{S}^2)$ $(= -tr(\boldsymbol{R}^2))$ but contain a large range of true $g_n$ values, such as the freestream region labelled (4). This shows that the standard two inputs can be insufficient in mapping to true $g_n$ accurately. NUM can also be found where the inputs and outputs vary in different directions. For example, some regions contain $tr(\boldsymbol{S}^2)$ and $tr(\boldsymbol{R}^2)$ varying in the $x_2$ direction, while true $g_n$ varies in the $x_1$ direction, such as those labelled (5). These causes of NUM justify why $n_{Cl}$ along the pure shear diagonal is higher compared to the rest of the $\boldsymbol{x}_{tr}$ space in **figure 6(g)-(i)**. As the majority of the white hatching contains true $g_1 \le 0$, its $g_n$ targets would mostly be unaffected by the *a posteriori* caveat, so the NUM there must be addressed.

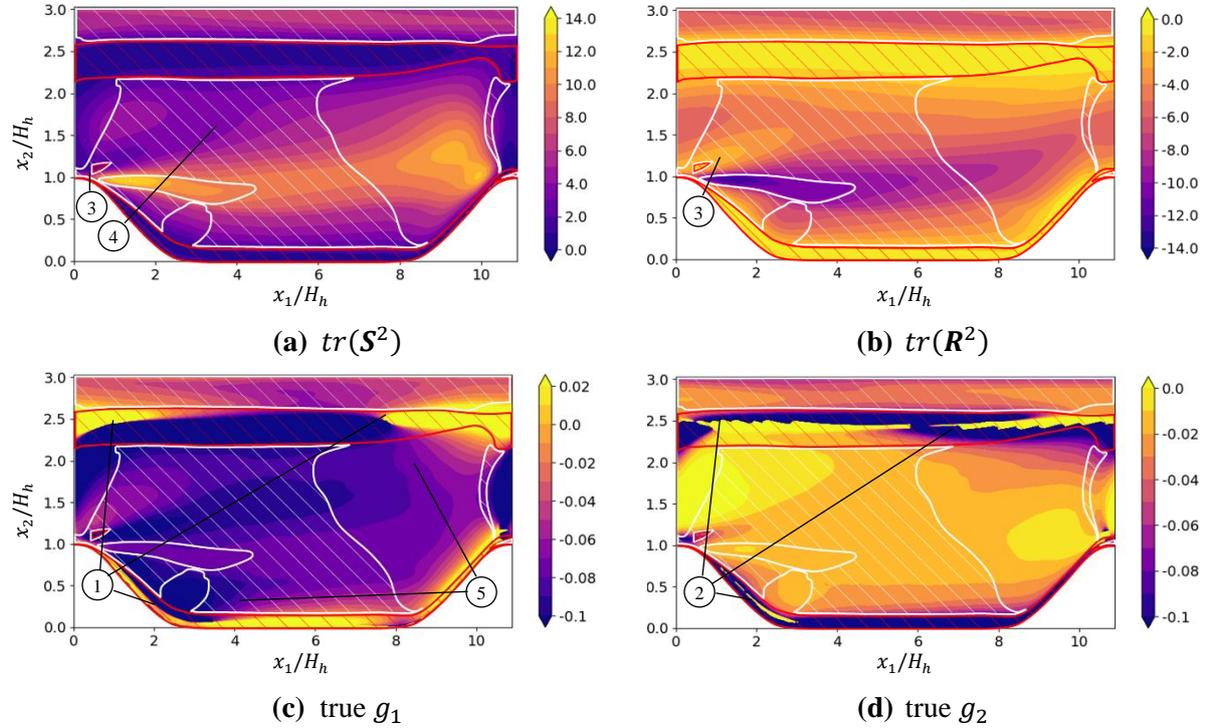

**(a)** $tr(\boldsymbol{S}^2)$

**(b)** $tr(\boldsymbol{R}^2)$

**(c)** true $g_1$

**(d)** true $g_2$

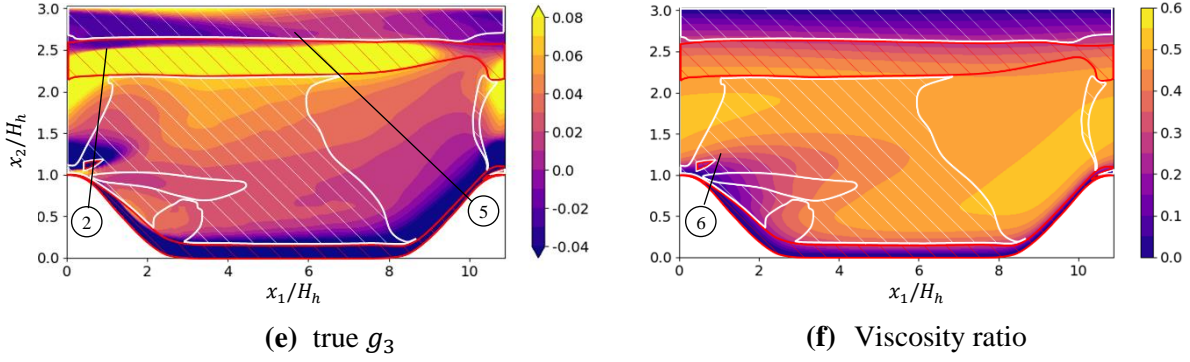

**(e)** true $g_3$            **(f)** Viscosity ratio

**Figure 7** Contour plots of **(a)-(b)** $\boldsymbol{x}_{tr}$ inputs, **(c)-(e)** true $g_n$, and **(f)** viscosity ratio $r_\nu$ for the periodic hills case. Red hatching shows regions of low strain and rotation flow where $tr(\boldsymbol{S}^2) < 2$ and $tr(\boldsymbol{R}^2) > -2$. White hatching shows regions of near pure shear flow that were not included in the red hatching and satisfy $-1 < tr(\boldsymbol{S}^2) + tr(\boldsymbol{R}^2) < 1$, $tr(\boldsymbol{S}^2) < 10$ and $tr(\boldsymbol{R}^2) > -10$.

### 4.4 Non-Correlation Between True $g_n$ and True $b_{ij}$

One may speculate whether the mapping between $\boldsymbol{x}_{tr}$ and true $g_n$ is the same as, or similar to between $\boldsymbol{x}_{tr}$ and true $b_{ij}$. If so, using true $b_{ij}$ instead of true $g_n$ as the NUM analysis outputs may be simpler and acceptable if most of the data points do not cause false positive and false negative diagnoses of NUM detailed in **Section 3.1**. **Figure 8** shows scatter plots of $\boldsymbol{x}_{tr}$ plotted against true $b_{ij}$ components from the periodic hills data. By qualitatively comparing **figure 8** with **figure 6(a)-(c)**, it can be seen that the mappings are not similar. For quantitative validation, Spearman's rank correlation coefficients (SRCCs) between true $g_n$ and true $b_{ij}$ were calculated (Corder & Foreman [2014](#)). The SRCC between two variables is a nonparametric measure of the statistical dependence between their ranked values, thereby indicating their monotonicity with a scale from -1 to 1, in which a SRCC of 1 or -1 indicates that the relationship is perfectly monotonic (Hollander *et al.* [2014](#)). The SRCC values in **table 1** demonstrate that the relationships between the true $g_n$ coefficients and true $b_{ij}$ components are not monotonic, hence the relationships in $\boldsymbol{x}_{tr}$ vs. true $g_n$ must be different to those in $\boldsymbol{x}_{tr}$ vs true $b_{ij}$. Therefore, true $g_n$ must not be substituted by true $b_{ij}$ in NUM analysis of 2D flow data. **Figure 8** also shows that $\boldsymbol{x}_{tr}$ has one-to-many relations with true $b_{ij}$, as the data points in each subplot may be grouped according to their true $b_{ij}$ values, and the groups may overlap in some parts of the $\boldsymbol{x}_{tr}$ space.

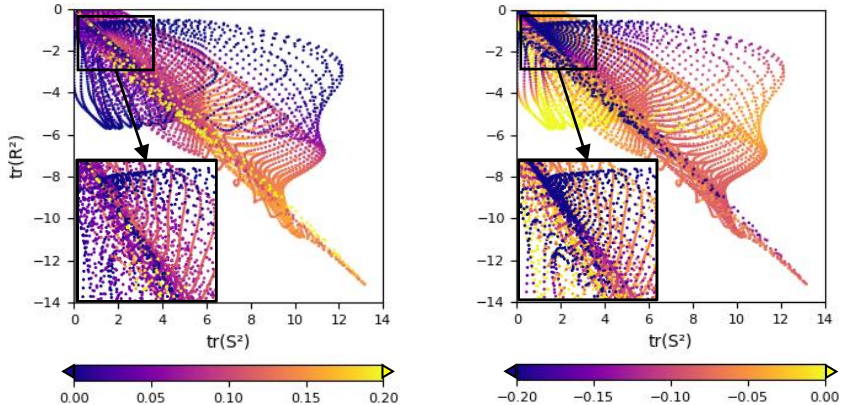

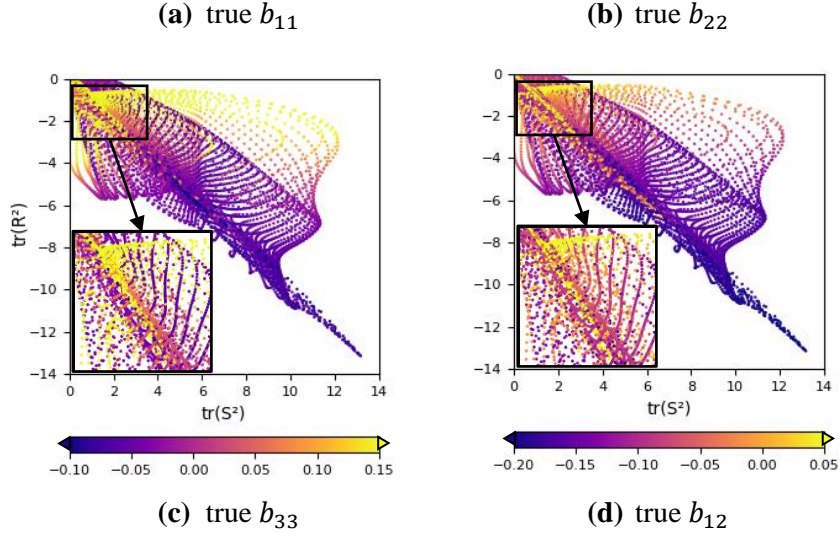

**Figure 8** Scatter plots of inputs $\boldsymbol{x}_{tr}$ vs. **(a)** true $b_{11}$, **(b)** true $b_{22}$, **(c)** true $b_{33}$, and **(d)** true $b_{12}$ components for the flow over periodic hills data.

**Table 1** Spearman's rank correlation coefficients between true $g_n$ coefficients and true $b_{ij}$ components

|            | true $b_{11}$ | true $b_{22}$ | true $b_{33}$ | true $b_{12}$ |
|------------|------|-------|-------|-------|
| true $g_1$ | 0.08  | -0.54 | 0.42  | 0.52  |
| true $g_2$ | -0.24 | 0.66  | -0.34 | -0.46 |
| true $g_3$ | 0.30  | 0.36  | -0.83 | -0.17 |

## 5 Impinging Jet Case Results

### 5.1 Non-Unique Mapping Quantification

The NUM quantification results for the impinging jet case are shown in **figure 9**. The scatter plots in **figure 9(a)-(c)** show that most of the points are located in the LSR subset where $tr(\boldsymbol{S}^2) < 2$ and $tr(\boldsymbol{R}^2) > -2$, as well as near the pure shear line. Similar to the periodic hills data, overlapping groups can be observed in these parts of the $\boldsymbol{x}_{tr}$ space for all true $g_n$ coefficients. For example, **figure 9(a)** shows that true $g_1 \approx -0.08$ or $-0.02$ for any value of $\boldsymbol{x}_{tr}$ along the pure shear line in the zoomed-in view, thus demonstrating one-to-many relations between $\boldsymbol{x}_{tr}$ and true $g_1$. Similar observations along the pure shear line can be made with true $g_2$ and $g_3$ in **figure 9(b)** and **figure 9(c)**, respectively.

20 000 data points were randomly selected for the clustering process due to memory constraints. Although this amounts to only 18% of all data points in **figure 9(a)**, a comparison with **figure 9(d)** shows that the selected points were still able to capture the overall distribution in the $\boldsymbol{x}_{tr}$ space, and thereby represent it. **Figure 9(d)-(f)** shows that the points have been reasonably grouped according to their true $g_n$ ($n \in \{1, 2, 3\}$) values by the clustering algorithm. For example, a comparison between **figure 9(b)** and **figure 9(e)** shows that the points along the pure shear diagonal that give true $g_2 \approx -0.02$ have been clustered in one group, true $g_2 \approx -0.06$ in another group and true $g_2 < -0.1$ have been allocated to a third group. These clusters enabled the $n_{CI}$ heat maps calculated with NNM shown in **figure 9(g)-(i)** to be made. Like the periodic hills results, the heat maps support qualitative NUM observations made around **figure 9(a)-(c)**, as $n_{CI}$ is mostly concentrated in LSR and near the pure shear line in the $\boldsymbol{x}_{tr}$ space: 89%, 87% and 90% of the total $n_{CI}$ in the true $g_1$, $g_2$, and $g_3$ heat maps, respectively, are found in the heat map grid cells along the pure shear line. Furthermore, the heat map results are supported by **figure 17(d)-(f)**, which show that running **Algorithm 1** with PBM instead gives similar $n_{CI}$ distribution trends.

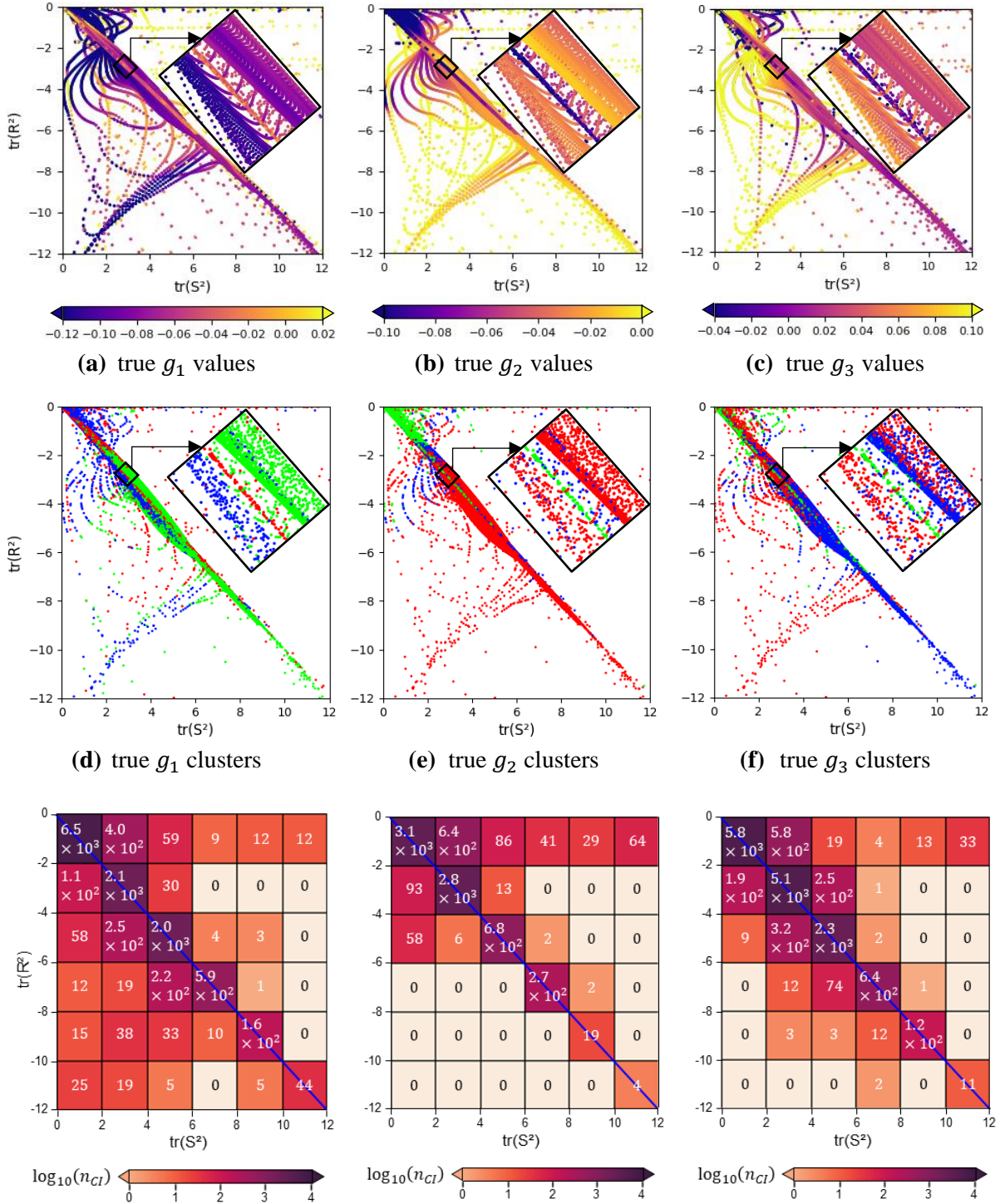

**(a)** true $g_1$ values     **(b)** true $g_2$ values     **(c)** true $g_3$ values

**(d)** true $g_1$ clusters     **(e)** true $g_2$ clusters     **(f)** true $g_3$ clusters

**(g)** heat map of $n_{CI}$ for true $g_1$     **(h)** heat map of $n_{CI}$ for true $g_2$     **(i)** heat map of $n_{CI}$ for true $g_3$

**Figure 9** NUM results for the impinging jet flow data. Subplots **(a)-(c)** show scatter plots of inputs $x_{tr}$ vs. outputs true $g_n$, **(d)-(f)** show clusters represented by colours: red, blue, and green for 20 000 random points, and **(g)-(i)** show heat maps of $n_{CI}$ given by NNM ($n = 3$), with the blue line indicating pure shear flow.

## 5.2 Low Strain and Rotation Regions

To identify regions in the domain that contain LSR and near pure shear flow, contours of the NUM analysis inputs and outputs are shown in **figure 10(a)-(b)** and **figure 10(c)-(e)**, respectively. The same hatching constraints used for the periodic hills data were applied. Two regions were found to contain LSR where $tr(S^2) < 2$ and $tr(R^2) > -2$ as indicated by red hatching: one near the top wall and another near the bottom wall. Similar to the periodic hills case, these regions contain $\partial \bar{u}_1 / \partial x_2$ inflection points, and satisfy Eq. (4.1) due to low values of $\partial \bar{u}_1 / \partial x_2$ compared to the rest of the domain. The mean streamwise velocity $\bar{u}_1$ profiles in the $x_2$ direction at the recirculation zones are flatter near the top wall compared to near the bottom wall. Therefore, low values of $\partial \bar{u}_1 / \partial x_2$ are sustained for a greater $x_2$ distance near the top wall, which lead to greater $x_2$ thickness in the red hatched region there (Jaramillo *et al.* 2012).

**Figure 10(c)-(e)** show both red hatched regions contain discontinuities in true $g_n$ with examples labelled (1). The cause of these discontinuities is the same as those in the periodic hills case – namely, very low $S_{11}$ and $S_{12}$ magnitudes. Although the discontinuities would result in NUM between $x_{tr}$ and true $g_n$, it has been explained in **Section 4.2** that the discontinuities are eliminated upon substitution in Eq. (2.6), and can therefore be ignored.

## 5.3 Near Pure Shear Regions

**Figure 10** shows that near pure shear flow satisfying Eq. (4.2) marked by white hatching dominates the domain. Further indication of this is given by comparing **figure 10(a)** with **(b)**, which demonstrates that the distributions of $tr(S^2)$ and $tr(R^2)$ are almost perfectly negatively correlated with each other. It is found that the white hatched regions contain absolute values of $S_{11}$ that are less than 0.5 and one order of magnitude smaller than $S_{12}$ and $R_{12}$. Therefore, $S_{12}$ and $R_{12}$ must have similar absolute values to satisfy the condition in Eq. (4.2), which is only possible if $|\partial \bar{u}_1 / \partial x_2| \gg |\partial \bar{u}_2 / \partial x_1|$ as shown in Eq. (2.6). It was observed that $|\partial \bar{u}_1 / \partial x_2|$ is two orders of magnitude greater than $|\partial \bar{u}_2 / \partial x_1|$ in these regions. This shows that $\partial \bar{u}_1 / \partial x_2$ is the only non-negligible velocity gradient in these regions of this case and therefore, inputs $tr(S^2)$ and $tr(R^2)$ can be approximated as:

$$tr(S^2) \approx \frac{k^2}{2\varepsilon^2} \left( \frac{\partial \bar{u}_1}{\partial x_2} \right)^2$$
$$tr(R^2) \approx -\frac{k^2}{2\varepsilon^2} \left( \frac{\partial \bar{u}_1}{\partial x_2} \right)^2$$

(5.1)

Eq. (5.1) supports the almost perfect negative correlation found between $tr(\boldsymbol{S}^2)$ and $tr(\boldsymbol{R}^2)$ in the white hatched regions. Moreover, its derivation gives insight into how near pure shear flow arises in the impinging jet case. **Figure 3(b)** shows various flow physics including recirculation, wall jets and their interface are contained in the white hatched regions and satisfy Eq. (5.1).

It was discussed in the periodic hills results that NUM exists where there are almost constant values of $\boldsymbol{x}_{tr}$ and a large range of true $g_n$. **Figure 10** shows that this occurs in the white hatched regions with examples labelled (2). Hence, the standard inputs $\boldsymbol{x}_{tr}$ are insufficient in mapping to true $g_n$ accurately in this case as well. Furthermore, **figure 10(a)-(b)** show that $tr(\boldsymbol{S}^2)$ and $tr(\boldsymbol{R}^2)$ vary in both $x_1$ and $x_2$ directions in the white hatched regions, resulting in the magnitude of the inputs increasing towards the free jet. Contrastingly, **figure 10(c)-(e)** show that true $g_n$ vary primarily in the $x_2$ direction. For example, the magnitude of true $g_1$ increases towards the horizontal centreline where $x_2/B = 2$, and the magnitude of true $g_2$ and $g_3$ approach zero. Therefore, in the outer parts of the white hatched regions such as where $2 < tr(\boldsymbol{S}^2) < 4$ and $-2 > tr(\boldsymbol{R}^2) > -4$ labelled (3), one-to-many relations (and therefore NUM) exist between $\boldsymbol{x}_{tr}$ and true $g_n$. This justifies why $n_{CI}$ increases as the magnitude of the inputs decrease along the pure shear line as shown in **figure 9(g)-(i)**. Similar to the periodic hills case, the NUM in the white hatched regions must be addressed as it only contains true $g_1 \leq 0$.

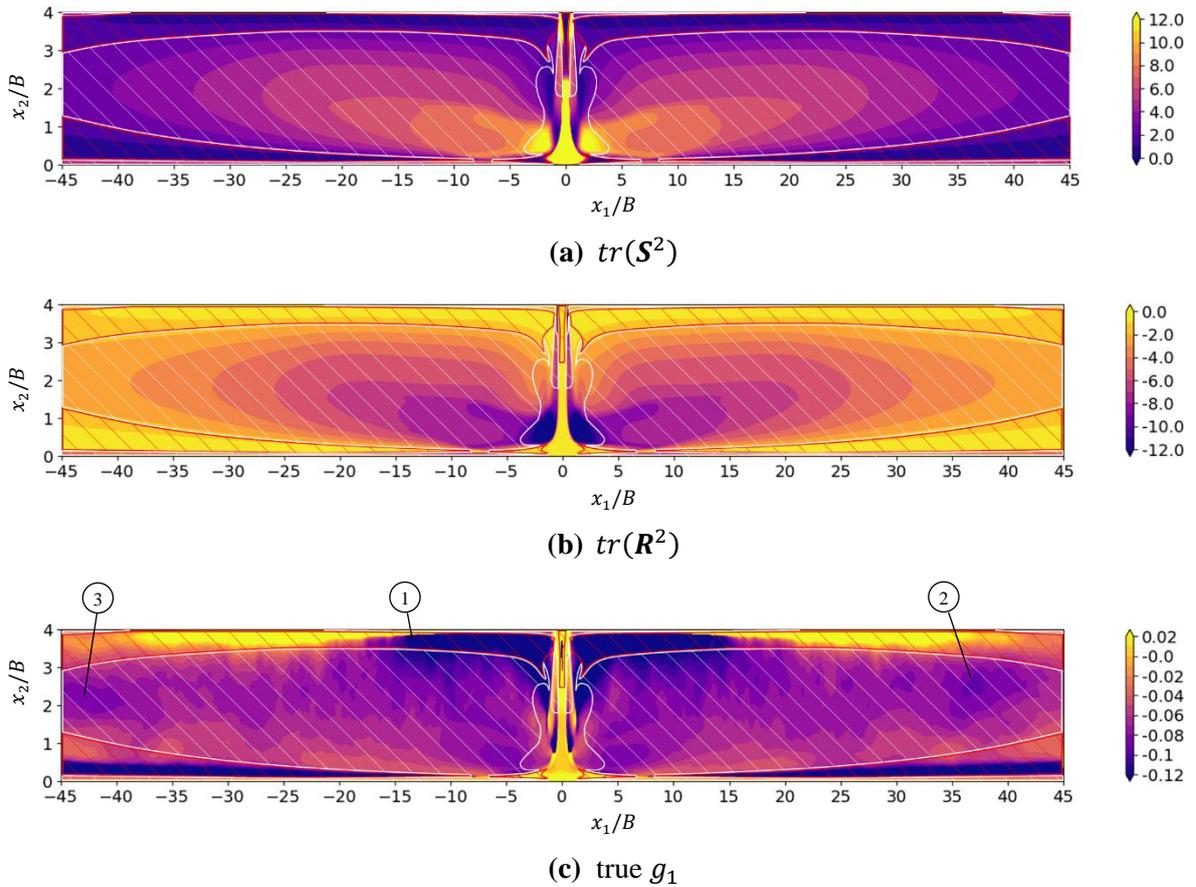

**(a)** $tr(\boldsymbol{S}^2)$

**(b)** $tr(\boldsymbol{R}^2)$

**(c)** true $g_1$

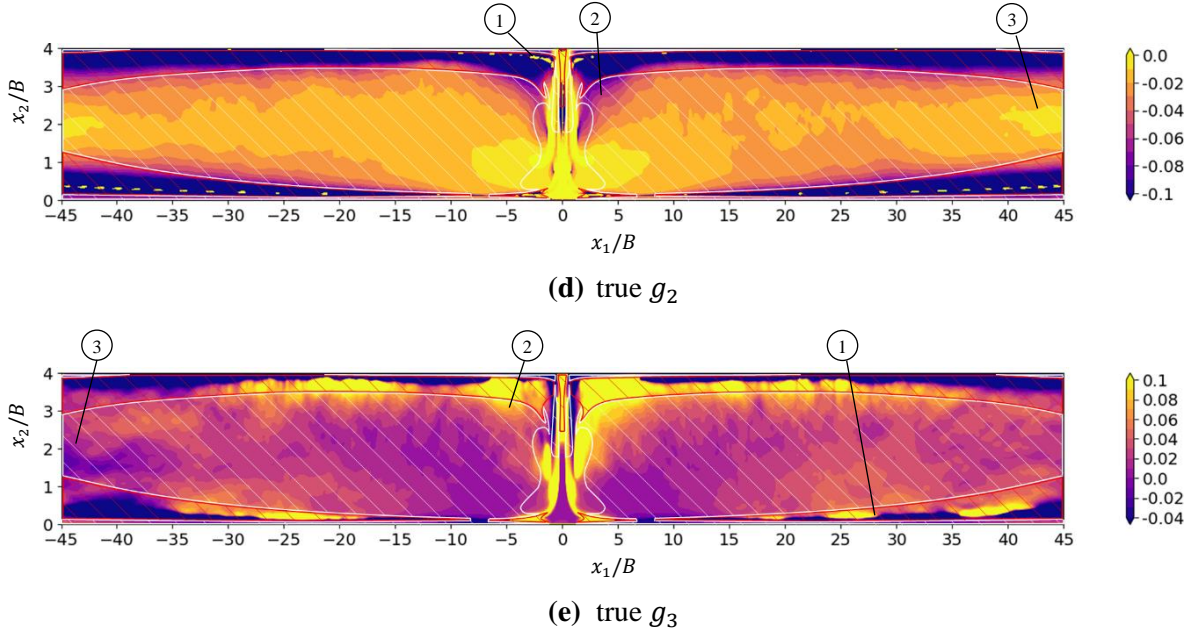

**(d)** true $g_2$

**(e)** true $g_3$

**Figure 10** Contour plots of the **(a)-(b)** $\boldsymbol{x}_{tr}$ inputs and **(c)-(e)** true $g_n$ for the impinging jet case. Red hatching shows regions of low strain and rotation flow where $tr(\boldsymbol{S}^2) < 2$ and $tr(\boldsymbol{R}^2) > -2$. White hatching shows regions of near pure shear flow that were not included in the red hatching and satisfy $-1 < tr(\boldsymbol{S}^2) + tr(\boldsymbol{R}^2) < 1$, $tr(\boldsymbol{S}^2) < 10$ and $tr(\boldsymbol{R}^2) > -10$.

## 6 Non-Unique Mapping Reduction

### 6.1 Viscosity Ratio $r_v$ as a Supplementary Input Variable

Having demonstrated that data from 2D flows can cause NUM in the GEVH, developing methods to reduce or, better still, eliminate NUM becomes a necessary step to improve TBML model training convergence and predictive accuracy. Previous studies showed that including a supplementary input variable (SIV) enabled NUM from 1D channel flow data to be eliminated, and SIVs have already been included in some TBML models in the literature (Kaandorp & Dwight [2020](#); Liu *et al.* [2021](#), Ellis & Xia [2023](#)). Therefore, the hypothesis that SIVs reduce or eliminate NUM caused by 2D flow data was tested. The tasks became choosing an SIV that could address NUM and demonstrating the improvement.

While one-to-many relations were found to be mostly concentrated in the LSR and near pure shear flow regions in both cases, **Sections 4.2** and **5.2** explained that those in LSR regions can be ignored. Therefore, the SIV was chosen to primarily address NUM in near pure shear flow, which can be attributed to causing the most $n_{CI}$ anyway – approximately 45% of total $n_{CI}$ in both flow cases. In this study, the SIV was chosen to be tailored for the periodic hills case. It was explained in **Section 4.3** that the standard $\boldsymbol{x}_{tr}$ inputs are almost constant but the values of true $g_n$ vary greatly near its separation point as shown in **figure 7**. As the separated flow is the defining flow feature, an SIV that can address NUM there was sought after, and it was believed that this could be achieved if the distribution of the SIV is similar to the true $g_n$ coefficients there. While different SIVs proposed in the literature were investigated as shown in **Appendix F**, only two qualified with their distributions: the ratio of convection to production of TKE, and the turbulence Reynolds number $Re_t$ that was used in the 1D channel flow studies. As the latter is simpler and its effectiveness in addressing 1D channel flow NUM has already been recognised, a variant of it which is more often found in data-driven turbulence models including TBML models called the viscosity ratio $r_\nu = \nu_t/(100\nu + \nu_t)$ was chosen as the SIV, where $\nu_t$ and $\nu$ are the eddy and kinematic viscosity, respectively (Volpiani *et al.* 2021; Wu *et al.* 2022; Ellis & Xia 2023). Eddy viscosity was calculated as $\nu_t = C_\mu k^2/\varepsilon$, where $C_\mu$ is a constant equal to 0.09. **Figure 7(f)** shows $r_\nu$ in the periodic hills case, which highly varies near the separation point in a similar manner to true $g_n$ as labelled (6). While $r_\nu$ was chosen as the SIV, the authors do not rule out well-established or new SIVs in the literature outperforming $r_\nu$ in reducing NUM. An investigation on comparing the performance of different SIVs would provide insightful results and may be subject to future work.

Scatter plots of true $g_n$ with $\boldsymbol{x}_{tr}$ on the $x$- and $y$-axes, and $r_\nu$ on the $z$-axis for the periodic hills data are shown in **figure 11(a)-(c)**. By comparing these to **figure 6(a)-(c)**, it can be seen that including $r_\nu$ enables more distinct true $g_n$ distributions, such that the points are more separated but closer to others with similar true $g_n$ values, which improves the smoothness of the mapping between inputs and true $g_n$. This can be clearly observed along the pure shear line of $tr(\boldsymbol{S}^2) = -tr(\boldsymbol{R}^2)$, thus qualitatively demonstrating that including $r_\nu$ as an SIV should reduce NUM in near pure shear flow. To facilitate quantitative investigation of the NUM reduction, the clustering process detailed in **Section 3.4** was extended to three dimensions. The modifications for this are detailed in **Appendix A**. For consistency with the results obtained from inputs $\boldsymbol{x}_{tr}$ in **figure 6**, all true $g_n$ coefficients were grouped into three clusters, which are shown in **figure 11(d)-(f)**. A comparison with **figure 11(a)-(c)** shows that the points have been reasonably clustered in accordance with their true $g_n$ ($n \in \{1, 2, 3\}$) values, and the clusters collapsed in the direction of the $r_\nu$ axis resemble those in **figure 6(d)-(f)**.

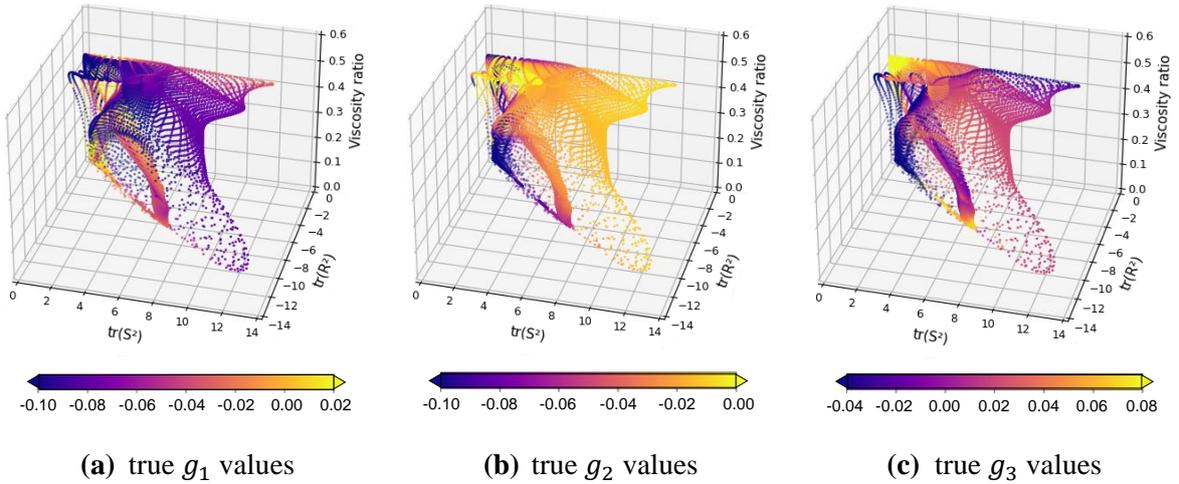

**(a)** true $g_1$ values      **(b)** true $g_2$ values      **(c)** true $g_3$ values

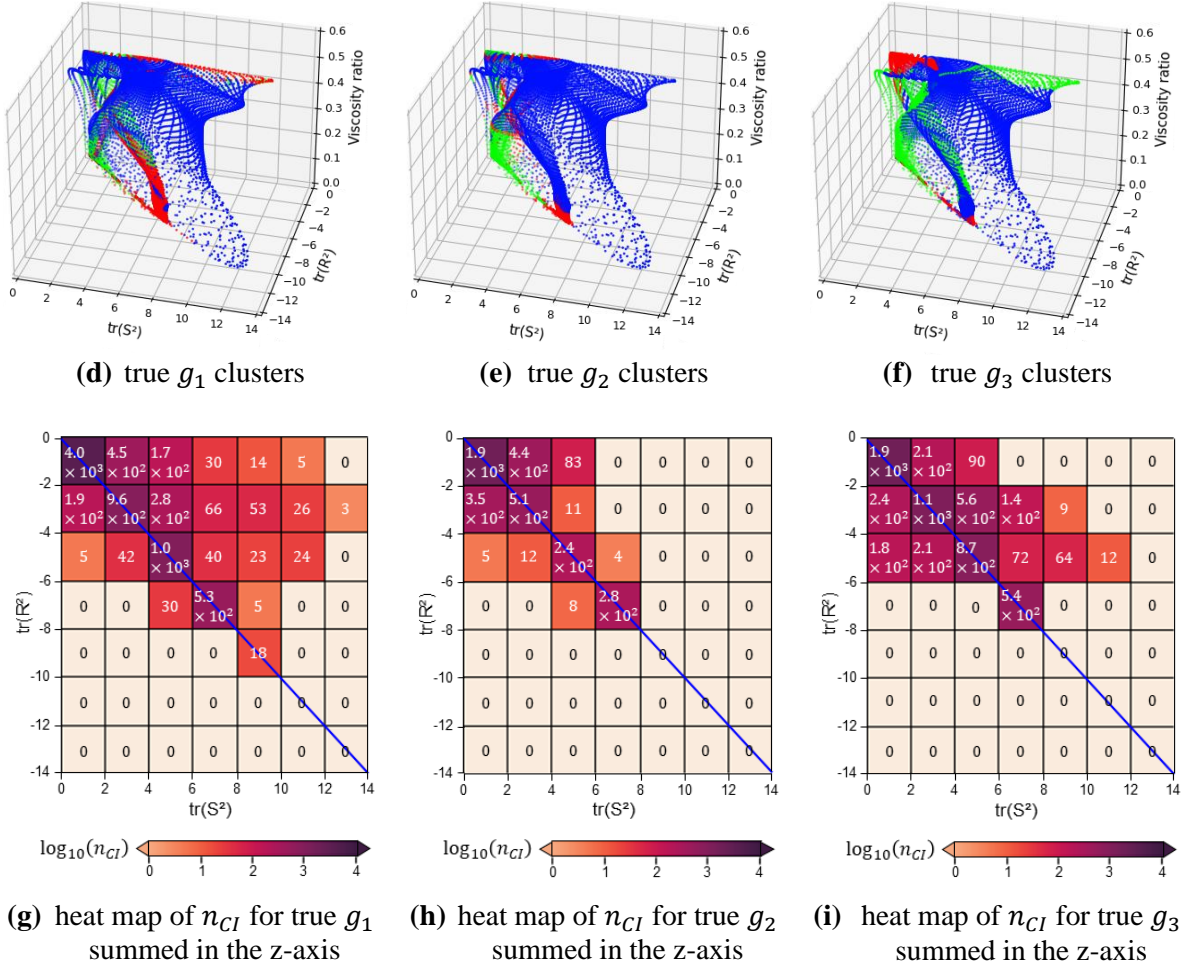

**(d)** true $g_1$ clusters    **(e)** true $g_2$ clusters    **(f)** true $g_3$ clusters

**(g)** heat map of $n_{CI}$ for true $g_1$ summed in the z-axis

**(h)** heat map of $n_{CI}$ for true $g_2$ summed in the z-axis

**(i)** heat map of $n_{CI}$ for true $g_3$ summed in the z-axis

**Figure 11** NUM results with inputs $\{\boldsymbol{x}_{tr}, r_v\}$ for the flow over periodic hills data. Subplots **(a)-(c)** show scatter plots of inputs $\boldsymbol{x}_{tr}$ and viscosity ratio $r_v$ vs. outputs true $g_n$, **(d)-(f)** show clusters represented by colours: red, blue, and green, and **(g)-(i)** show summed heat maps of $n_{CI}$ given by NNM ($n = 3$), with the blue line indicating pure shear flow.

## 6.2    Non-Unique Mapping Reduction Quantification

To quantify the NUM, the $\{\boldsymbol{x}_{tr}, r_v\}$ input space was partitioned into a 3D grid, and the NUM quantification method detailed in **Section 3.5** was run with NNM extended to three dimensions, whereby Eq. (3.2) was modified as:

$$d = \sqrt{\left(tr(\boldsymbol{S}^2)_p - tr(\boldsymbol{S}^2)_q\right)^2 + \left(tr(\boldsymbol{R}^2)_p - tr(\boldsymbol{R}^2)_q\right)^2 + \left(r_{v_p} - r_{v_q}\right)^2} \qquad (6.1)$$

and 3D heat maps of $n_{CI}$ were generated. To compare these results with the 2D heat maps given by just inputs $\boldsymbol{x}_{tr}$ in **figure 6(g)-(i)** denoted HM1, the 3D heat maps were summed in the direction of the $r_v$ axis to give 2D heat maps of $n_{CI}$ in the $\boldsymbol{x}_{tr}$ space as shown in **figure 11(g)-(i)**. These were subtracted from HM1 to give difference arrays, which show the difference in $n_{CI}$ throughout the $\boldsymbol{x}_{tr}$ space after including $r_v$ as an SIV. The process of generating the difference arrays is depicted in **figure 12**.

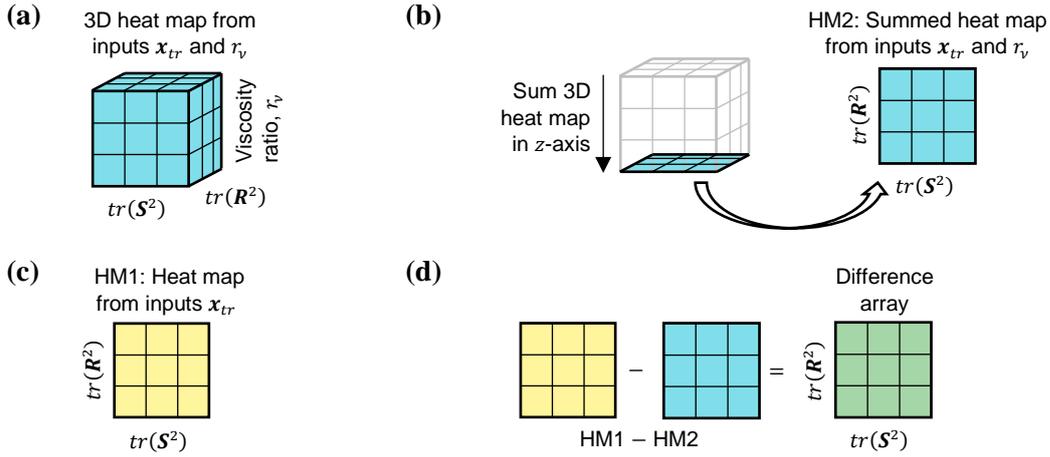

**Figure 12** Process of obtaining the difference arrays: **(a)** simplified schematic of a 3D heat map containing $n_{CI}$ for inputs $\boldsymbol{x}_{tr}$ and $r_v$, **(b)** summation of the 3D heat map in the $r_v$ axis to give the flattened heat map, **(c)** simplified schematic of the 2D heat map containing $n_{CI}$ for inputs $\boldsymbol{x}_{tr}$, **(d)** subtraction of the flattened heat map from the 2D heat map to give the difference array.

**Figure 13(a)-(c)** show the difference arrays for the periodic hills data. These results demonstrate that NUM can reduce in the vast majority of the domain when an SIV is included. Choosing $r_v$ as the SIV led to most improvement being located along the pure shear line as initially targeted. **Table 2** shows including $r_v$ reduced $n_{CI}$ by 45.7%, 60.3%, and 57.9% for true $g_1$, $g_2$, and $g_3$, respectively, in total. Difference arrays were also generated for the impinging jet data as shown in **figure 13(d)-(f)**. It can be observed that much of the improvement is again located along the pure shear line, thus $r_v$ as an SIV was able to target this subset of the $\boldsymbol{x}_{tr}$ input space effectively in this case too. However, large increases in $n_{CI}$ can also be observed – especially in the true $g_3$ difference array, which led to lower total $n_{CI}$ reductions of 44.5%, 34.2%, and 5.8% for true $g_1$, $g_2$, and $g_3$, respectively, in the $\boldsymbol{x}_{tr}$ space.

It is suspected that when the input space increases in dimension due to the inclusion of another input variable, the close points of a focal point belonging to the same cluster may possibly move further away from the focal point in the input space. This can cause other points in close proximity to be promoted as the focal point's close points. If these new close points do not belong to the same cluster as the focal point, then an increase in the number of conflicting instances occurs. Where the points move to in the new dimension is due to the choice of SIV. This indicates that the choice of SIV can lead to different amounts of NUM reduction in (i) different true $g_n$ coefficients and (ii) different flow case data. The NUM reduction in the periodic hills data may be significantly greater than in the impinging jet data due to $r_v$ being specifically chosen to reduce NUM in the periodic hills data near the separation point. Therefore, this shows that causes of NUM in 2D flow data used for TBML model training should be identified, and the choice of SIVs should be tailored towards the flow data. Nonetheless, reductions in NUM for both cases and all true $g_n$ coefficients show that including SIVs should be a generally viable approach to reduce NUM.

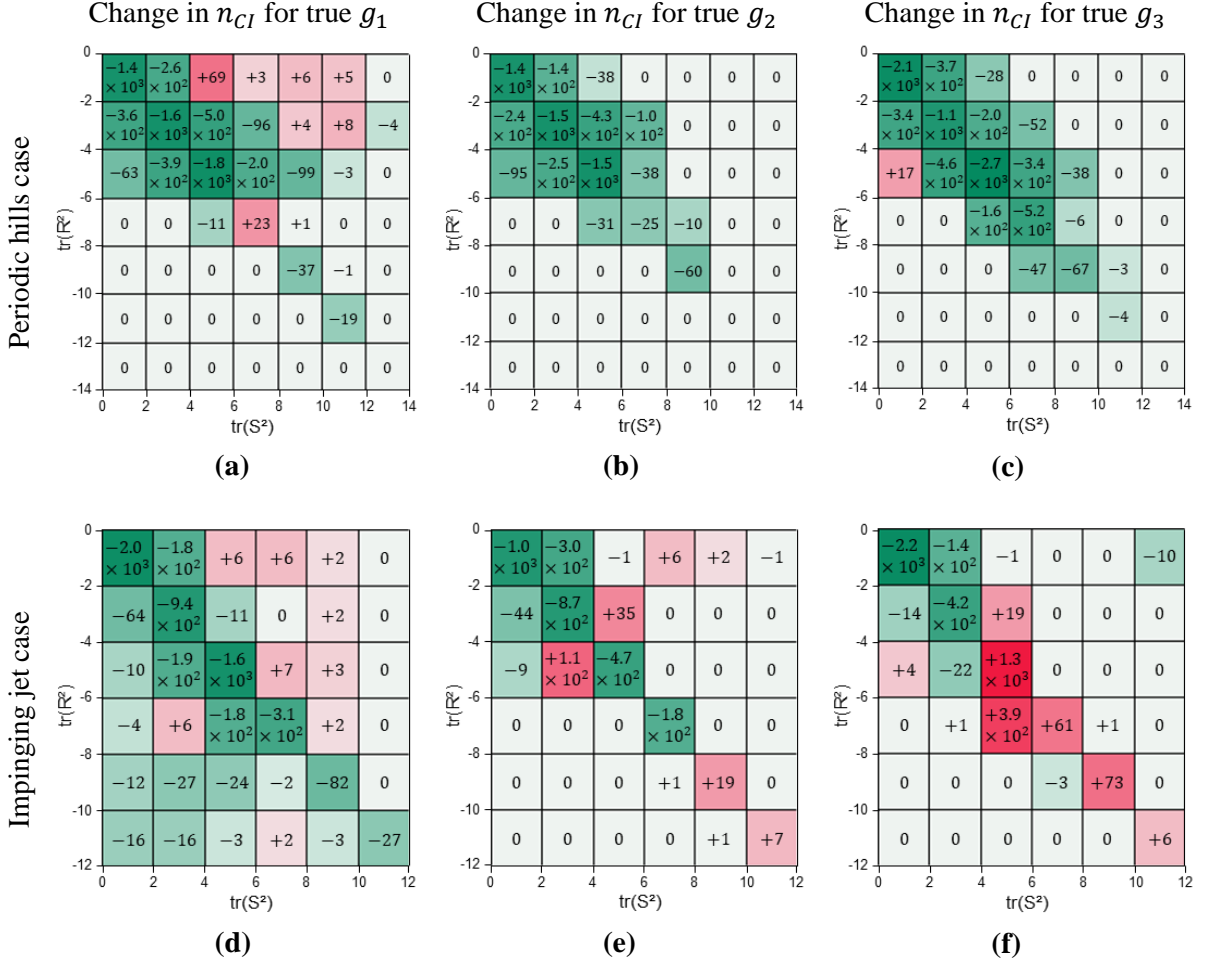

**Figure 13** Difference arrays showing the change in number of conflicting instances $\Delta n_{CI}$ throughout the $\boldsymbol{x}_{tr}$ space after including viscosity ratio $r_\nu$ as a supplementary input variable. For the periodic hill case, $\Delta n_{CI}$ for true $g_1$ is shown in (**a**), for true $g_2$ in (**b**), and for true $g_3$ in (**c**), while for the impinging jet case, $\Delta n_{CI}$ for true $g_1$ is shown in (**d**), for true $g_2$ in (**e**), and for true $g_3$ in (**f**).

**Table 2** Total percentage reduction in conflicting instances across entire $\boldsymbol{x}_{tr}$ space after including viscosity ratio as a supplementary input variable.

| Case | true $g_1$ | true $g_2$ | true $g_3$ |
|---|---|---|---|
| Periodic hills (%) | 45.7 | 60.3 | 57.9 |
| Impinging jet (%) | 44.5 | 34.2 | 5.8 |

### 6.3 Training Experiment

To demonstrate the effect of including $r_\nu$ as an SIV in TBML model training, two TBNNs were trained. The TBNNs were identical, except the inputs of the first model were the two invariants shown in Eq. (2.4), while the inputs of the second model were the two invariants and $r_\nu$. These models will be referred to as tbnn-1 and tbnn-2. The periodic hill case that has been analysed thus far in the paper with steepness factor $\beta = 1.5$ was used for training, whereas flow over a periodic hill with $\beta = 1$ and 1.2 were used for validation and testing, respectively. As the cases are 2D, the two-dimensional version of TBNN with three tensor basis was used (Pope 1975). The models had three hidden layers with 10 hidden nodes per layer and rectified linear unit activation. Training was undertaken with a learning rate of 0.001, batch size of 32 and the Adam optimizer. Mean squared error loss between predicted $b_{ij}$ and DNS $b_{ij}$ was used as the true $g_n$ coefficient values can tend towards infinity in some flow regions. The models were evaluated on the validation set after every two epochs. If the average validation error over the last three evaluations was higher than the average over the three before, then model training was stopped early.

As shear anisotropy governs the shear flow downstream of the hill crest and subsequent reattachment, contours of predicted $b_{12}$ are shown in **figure 14** (Xiao *et al.* 2020). The colorbar has been set to the lower and upper realizability bounds of -0.5 and 0.5. A comparison between the RANS prediction in **figure 14(a)** with the DNS prediction in **figure 14(d)** shows that while the distribution of the RANS prediction matches closely with the DNS, differences in magnitudes can be observed close to the wall, and above the hill crest. The tbnn-1 prediction does not match the DNS magnitudes closely either, due to discontinuities in the predictions which are believed to be caused by NUM. The tbnn-2 prediction contains magnitudes closer to the DNS compared to RANS, especially near the upper wall, above and downstream of the hill crest. Furthermore, it contains significantly less discontinuities compared to tbnn-1, hence suggesting a reduction in NUM can reduce discontinuous TBML model predictions.

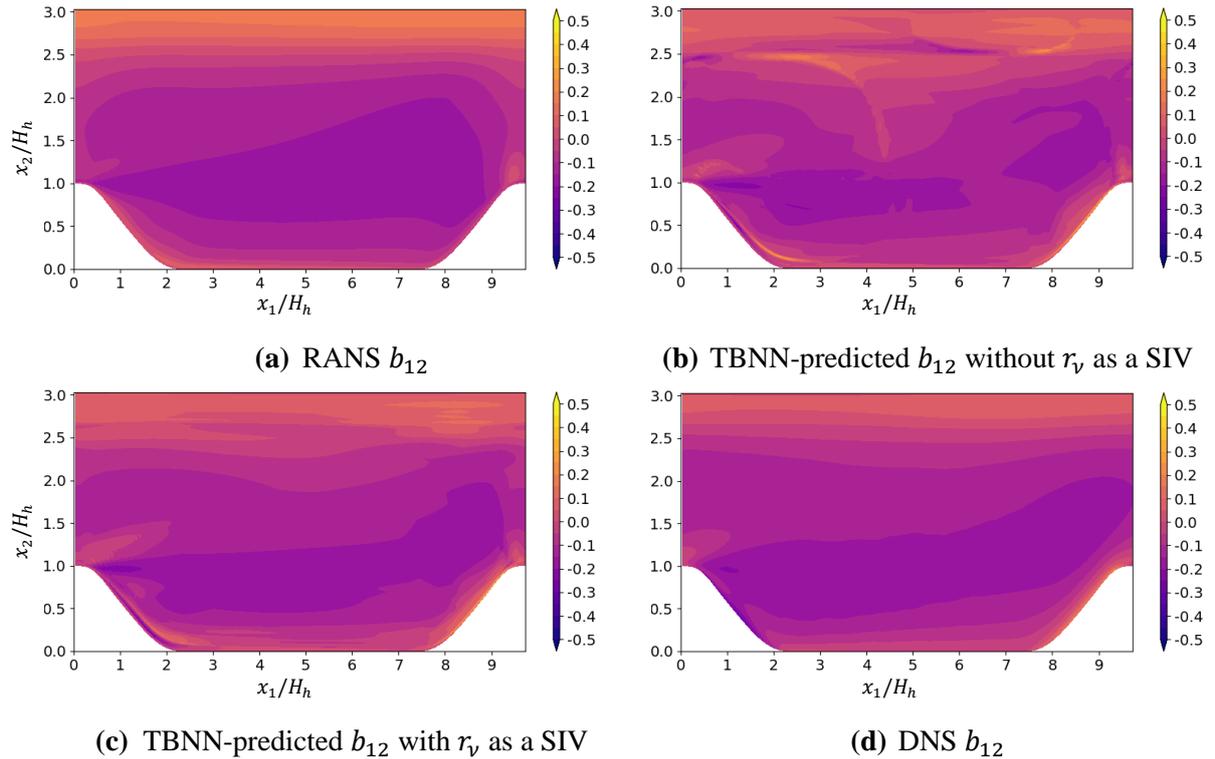

**(a)** RANS $b_{12}$

**(b)** TBNN-predicted $b_{12}$ without $r_\nu$ as a SIV

**(c)** TBNN-predicted $b_{12}$ with $r_\nu$ as a SIV

**(d)** DNS $b_{12}$

**Figure 14** Shear anisotropy $b_{12}$ predicted by **(a)** RANS SST by McConkey *et al.* (2021), **(b)** TBNN with the two standard invariant inputs: $tr(\mathbf{S}^2)$ and $tr(\mathbf{R}^2)$, **(c)** TBNN with the two standard invariant inputs and $r_\nu$ as a SIV, and **(d)** DNS by Xiao *et al.* (2020) for the $\beta$ = 1.2 periodic hill test case.

The mean squared error of the RANS, tbnn-1 and tbnn-2 predictions compared to DNS for the non-zero anisotropy components are given in **table 3**. These quantitative results show that for all components,

tbnn-1 gave more accurate predictions than RANS, and the inclusion of input $r_\nu$ in tbnn-2 improved the predictions further. For example, a 24% improvement in the accuracy of $b_{12}$ was observed between tbnn-1 and tbnn-2.

**Table 3** Mean squared error of predicted anisotropy components compared to true values from DNS for the $\beta = 1.2$ periodic hill test case

| Method | Anisotropy Component | | | |
|---|---|---|---|---|
| | $b_{11}$ | $b_{12}$ | $b_{22}$ | $b_{33}$ |
| RANS | 0.0305 | 0.0041 | 0.0277 | 0.0095 |
| TBNN without $r_\nu$ | 0.0065 | 0.0024 | 0.0067 | 0.0066 |
| TBNN with $r_\nu$ | 0.0044 | 0.0014 | 0.0054 | 0.0044 |

## 7    Conclusion

A method for quantifying non-unique mapping (NUM) in data-driven Reynolds averaged turbulence models was proposed and deployed on the general effective viscosity hypothesis, which forms the basis of many such models in the literature, such as tensor-basis neural networks. Using data from a flow over periodic hills case and an impinging jet case, this study confirms for the first time that NUM can exist in these models with the standard invariant inputs when 2D cases are used to train them. The magnitude of this finding is underscored by the fact that such flow cases have been used to train most of these models in the literature. Furthermore, this discovery was demonstrated on two very different flow cases, which show that NUM is not limited to certain flow physics in the training data.

To identify types of flows that cause NUM, a pre-training metric for quantifying NUM throughout the input space called conflicting instances $n_{CI}$ was proposed in the method. For both flow cases, $n_{CI}$ was found to be highly concentrated in regions containing (i) low strain and rotation (LSR), and (ii) near pure shear flow. The former is caused by discontinuities in the target outputs, and the latter is due to various causes of one-to-many relations between the invariant inputs and the target outputs. These regions combined were found to cause approximately 76% and 89% of total $n_{CI}$ in the periodic hills and impinging jet case data, respectively. While it was found that NUM caused by LSR can be ignored due to vanishing discontinuities in the models, near pure shear flow gave the highest $n_{CI}$ regardless – approximately 45% of total $n_{CI}$ in both cases. These consistent findings should urge the community to examine their training data from cases containing near pure shear flow to potentially reduce NUM significantly in their data-driven turbulence models. As LSR and/or near pure shear flow (including homogeneous shear flow) exists in many flow cases, these findings are widely applicable and can help develop more accurate models for cross-case training, which is a critical challenge in the field today.

The method was repeated with viscosity ratio as a supplementary input variable (SIV) which was chosen for reducing NUM from pure shear flow in the periodic hills case. This led to total $n_{CI}$ being reduced for all outputs in both different flow cases. Therefore, it is demonstrated for the first time that including SIV(s) is a viable approach for reducing NUM caused by 2D flows in these data-driven turbulence models. This result also demonstrates the merit in choosing viscosity ratio and the proposed solution for reducing NUM. However, the overall $n_{CI}$ reduction varied with different output coefficients and flow cases: 46-60% for the periodic hills data and 6-45% for the impinging jet data. Smaller reductions for the impinging jet case were partly due to increases in $n_{CI}$ for some outputs in certain regions of the input space, which are believed to be caused by the choice of SIV. Given that viscosity ratio was chosen specifically to reduce $n_{CI}$ in the periodic hills data from near the separation region, this shows that the input and target outputs of each training case should always be analysed, and the choice of SIV(s) should be tailored towards the flow physics of the cases in order to effectively reduce NUM. Including the SIV in training a tensor basis neural network was found to improve prediction accuracy for all anisotropy components. As the types of testing flow cases are unknown in model deployment, SIVs that effectively reduce NUM over different types of flows should be investigated in the interest of cross-case training.

## Acknowledgements

The authors would like to acknowledge the assistance given by Research IT and the use of the Computational Shared Facility at The University of Manchester.

## Funding

This work was supported by the UK Engineering and Physical Sciences Research Council (grant numbers EP/W033542/1 and EP/T012242/2).

## Author Declarations

The authors have no conflicts to disclose.

## Data Availability Statement

The data that support the findings of this study are available from the corresponding author upon reasonable request.

# Appendix A: Clustering Process

## A.1 Min-Max Normalization

While the clustering algorithm was required to group points of similar true $g_n$ ($n \in \{1, 2, 3\}$) values together, it was desirable to a lesser extent for it to also group points with similar $\boldsymbol{x}_{tr}$ (and $r_\nu$) values, as this led to points with similar inputs vs true $g_n$ ($n \in \{1, 2, 3\}$) relation to belong to the same cluster. As the flow cases had different ranges of $tr(\boldsymbol{S}^2)$, $tr(\boldsymbol{R}^2)$, $r_\nu$, and true $g_n$ values, min-max normalization was performed to ensure that the clustering would not be affected by their scales of magnitude (Everitt *et al.* 2011; Zheng & Casari 2018):

$$\tilde{x} = \frac{a(x - \min(x))}{\max(x) - \min(x)}$$

$$a = \begin{cases} 0.1, & x \in \{tr(\boldsymbol{S}^2), tr(\boldsymbol{R}^2), r_\nu\} \\ 1, & x \in \{\text{true } g_n\} \end{cases} \tag{A1}$$

where $x$ in this appendix is an arbitrary variable that may represent $tr(\boldsymbol{S}^2)$, $tr(\boldsymbol{R}^2)$, $r_\nu$, or true $g_n$ ($n \in \{1, 2, 3\}$), and $\tilde{x}$ is their min-max normalized value. The factor $a$ ensured the range of normalized true $g_n$ values was 10 times greater than the range of normalized input values, enabling the clustering algorithm to be more weighted towards grouping points with similar true $g_n$ values than similar $tr(\boldsymbol{S}^2)$, $tr(\boldsymbol{R}^2)$ or $r_\nu$ values (Rebala *et al.* 2019).

## A.2 Complete Linkage Agglomerative Clustering

The normalized data points were clustered using complete linkage agglomerative (CLA) clustering. Agglomerative clustering algorithms begin by treating each point as a cluster of its own. The two clusters that are separated by the closest distance are combined, and this step is repeated until a pre-specified number of clusters remain (Everitt *et al.* 2011). In CLA, the distance between two clusters is given by the distance between their furthest points. The Euclidean distance was used with inputs $\boldsymbol{x}_{tr}$:

$$d(p, q) = \sqrt{\left(\widetilde{tr(\boldsymbol{S}^2)}_p - \widetilde{tr(\boldsymbol{S}^2)}_q\right)^2 + \left(\widetilde{tr(\boldsymbol{R}^2)}_p - \widetilde{tr(\boldsymbol{R}^2)}_q\right)^2 + \left(\widetilde{g_{n_p}^t} - \widetilde{g_{n_q}^t}\right)^2} \tag{A2}$$

and with inputs $\boldsymbol{x}_{tr}$ and $r_\nu$:

$$d(p, q) = \sqrt{\left(\widetilde{tr(\boldsymbol{S}^2)}_p - \widetilde{tr(\boldsymbol{S}^2)}_q\right)^2 + \left(\widetilde{tr(\boldsymbol{R}^2)}_p - \widetilde{tr(\boldsymbol{R}^2)}_q\right)^2 + \left(\widetilde{r_{\nu_p}} - \widetilde{r_{\nu_q}}\right)^2 + \left(\widetilde{g_{n_p}^t} - \widetilde{g_{n_q}^t}\right)^2} \tag{A3}$$

where $d$ is the Euclidean distance between two arbitrary data points $p$ and $q$, and true $g_n$ is represented by $g_n^t$. The tilde overbar $\widetilde{\square}$ denotes a min-max normalized quantity from Eq. (A1), and subscripts $p$ and $q$ indicate which data point the quantity is from.

## Appendix B: Channel Flow Analysis

**Figure 15(a)** shows the nondimensional velocity $u^+$ and shear velocity gradient $du^+/dy^+$ profiles for the channel flow case described in **figure 2**. It is clear that the velocity follows the well-established law of the wall profile given by equations $u^+ = y^+$ and $u^+ = \ln(y^+)/\kappa + C^+$, where $\kappa = 0.41$ and $C^+ = 5.5$. **Figure 15(b)** shows profiles of the inputs and true outputs of a TBML model for 1D channel flow without SIV. Non-dimensional shear velocity gradient $\alpha$ is the only input as explained in **Section 2.3**, and true $g_n$ ($n \in \{1, 2, 3\}$) are the outputs. Drawing horizontal lines across **figure 15(b)** shows that $\alpha$ has a one-to-many relation with the true outputs.

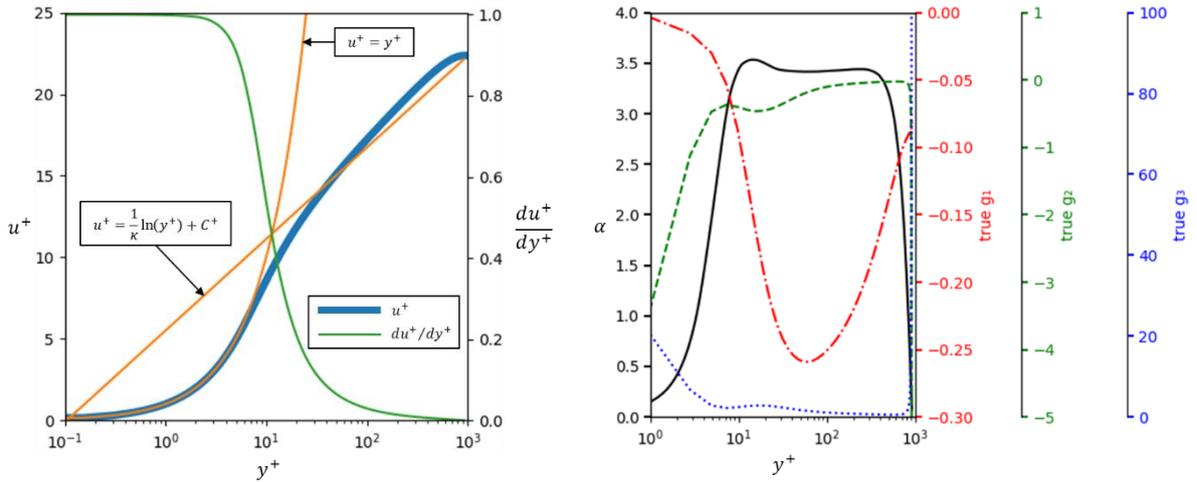

**(a)** Velocity $u^+$ and shear velocity gradient $du^+/dy^+$ profiles of the channel flow case given by RANS SST. The velocity and distance from the wall are nondimensionalised as $u^+ = \bar{u}_1/u_\tau$ and $y^+ = x_2 u_\tau/\nu$, where friction velocity $u_\tau = \sqrt{\tau_w/\rho}$ and $\tau_w$ is the wall shear stress.

**(b)** Non-dimensional shear velocity gradient $\alpha$ ($= (k/\varepsilon)(\partial \bar{u}_1/\partial x_2)$) given by RANS SST, true $g_1$, true $g_2$, and true $g_3$ profiles against nondimensional distance from the wall $y^+$.

**Figure 15** Profiles of **(a)** velocity and velocity gradient, and **(b)** input $\alpha$ and true $g_n$ ($n \in \{1, 2, 3\}$) for the 1D channel flow case.

**Figure 16** shows the NUM quantification results for the channel flow case. As $\alpha$ is the only input, the input space simplifies to one dimension which can be represented on a number line. This allowed cell-centre values of $\alpha$ to be plotted against true $g_n$ coefficients in **figure 16(a)-(c)**. Drawing vertical lines on these plots shows there are one-to-many relations between input $\alpha$ and outputs true $g_n$ coefficients. The clustering approach described in **Section 3.4** was used to cluster these points, and the allocated cluster of each point are indicated by their color – blue or red – in **figure 16(a)-(c)**. This representation allows clear visualisation and assessment on whether the clustering approach can identify the two different branches shown in **figure 2(a)**. **Figure 16(a)-(c)** show that the approach can approximately cluster the points based on the two branches. While the clustering approach did not identify the branches perfectly (as some clusters contain points from both branches), most points that share similar values of $\alpha$ but have significantly different true $g_n$ ($n \in \{1, 2, 3\}$) values were successfully allocated into different clusters. Therefore, we believe the current clustering approach gives acceptable performance. With these clusters, Euclidean distance $d$ was calculated as $d = \sqrt{\left(\alpha_p - \alpha_q\right)^2}$ instead of Eq. (3.2) to generate heat maps of $n_{CI}$ with the NNM described in **Section 3.5**. **Figure 16(d)-(f)** show that these heat maps effectively capture regions of the input space containing conflicting instances as indicators of NUM – especially where $3.25 < \alpha < 3.5$ in the true $g_2$ and true $g_3$ results which contains the log-law region. In future work, we believe the clustering algorithm may be improved by enabling neighbouring points in the physical flow domain to be easily clustered together, which would allow the two branches to be more easily identified and more accurate results of $n_{CI}$ can be calculated.

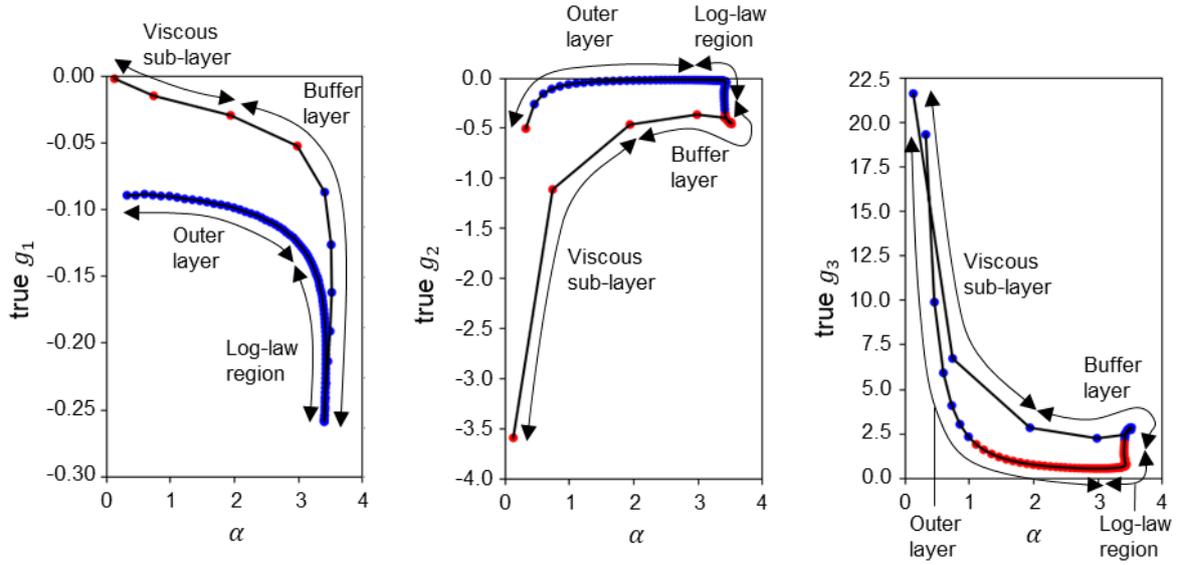

**(a)** true $g_1$ values and clusters   **(b)** true $g_2$ values and clusters   **(c)** true $g_3$ values and clusters

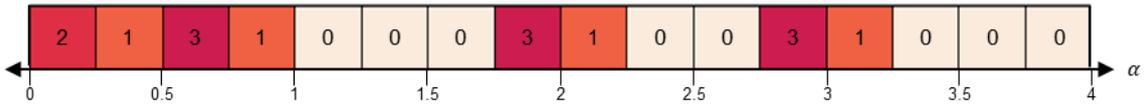

**(d)** heat map of $n_{CI}$ for true $g_1$

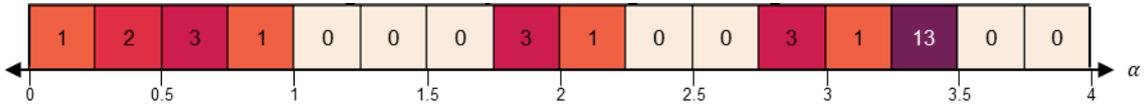

**(e)** heat map of $n_{CI}$ for true $g_2$

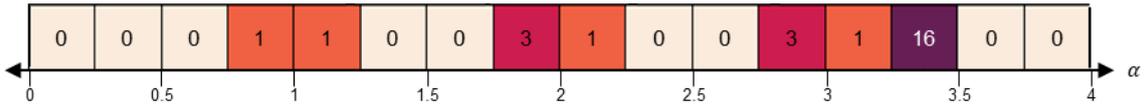

**(f)** heat map of $n_{CI}$ for true $g_3$

**Figure 16** NUM results for the channel flow data. Subplots **(a)-(c)** show scatter points of the input $\alpha$ vs. outputs true $g_n$ ($n \in \{1, 2, 3\}$) data colored in blue or red according to their cluster. Subplots **(d)-(f)** show heat maps of $n_{CI}$ given by NNM ($n = 2$).

## Appendix C: Proximity Box Method Results

**Figure 17** shows heat maps of $n_{CI}$ obtained by running **Algorithm 1** with PBM. The box length $l$ and height $h$ depicted in **figure 5(b)** were set to $1/8$ times the grid spacing for both cases, which allowed the box dimensions to scale with the grid spacing. Comparing **figure 6(g)-(i)** with **figure 17(a)-(c)**, and **figure 9(g)-(i)** with **figure 17(d)-(f)** shows that the heat maps obtained with NNM and PBM have very similar $n_{CI}$ distribution trends. Both methods show that $n_{CI}$ is highly concentrated in the LSR subset of $\boldsymbol{x}_{tr}$, and near the pure shear line where $tr(\boldsymbol{S}^2) = -tr(\boldsymbol{R}^2)$. Therefore, the results given by NNM have been validated with those from PBM. The NNM and PBM results only differ in the magnitude of $n_{CI}$, as the use of a proximity box in PBM naturally exacerbates $n_{CI}$ values in regions of the input space that contain densely packed points belonging to different clusters and reduces $n_{CI}$ where points are sparse.

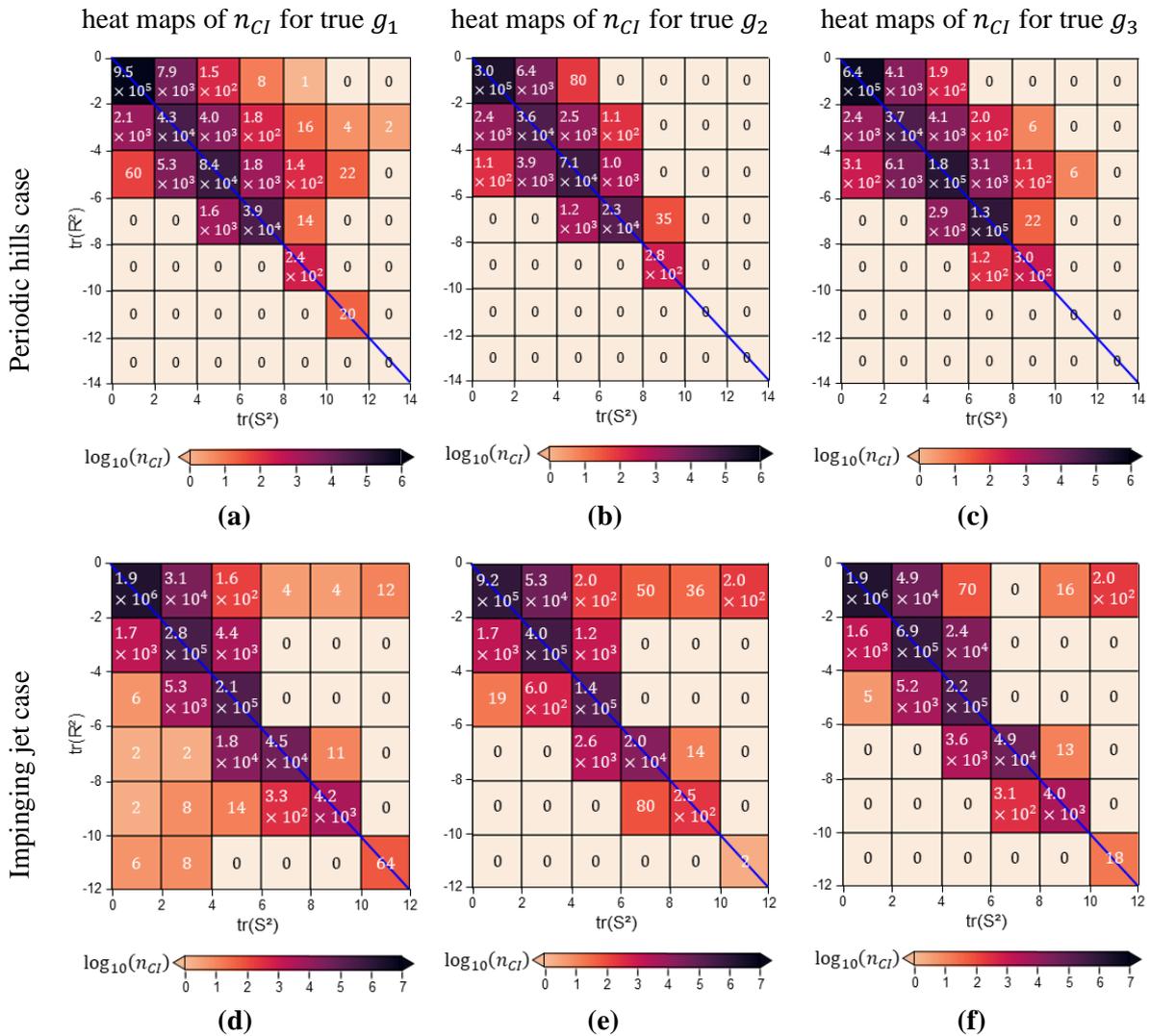

**Figure 17** Heat maps of $n_{CI}$ between inputs $\boldsymbol{x}_{tr}$ and outputs true $g_n$ given by PBM. The heat maps for the periodic hill case are shown in subfigures **(a)**, **(b)**, and **(c)**, while the heat maps for the impinging jet case are shown in subfigures **(d)**, **(e)**, and **(f)**.

## Appendix D: Limits of True $g_n$ Terms

The conventional $(\varepsilon, \delta)$ definition of a limit for a multivariable infinite limit can be written as follows: Let $f$ be a function of two variables $(x, y)$ whose domain $D$ includes points arbitrarily close to $(a, b)$. We say that $f$ tends to $\infty$ as $(x, y)$ approaches $(a, b)$, and we write:

$$\lim_{(x,y) \to (a,b)} f(x, y) = \infty \tag{C1}$$

if for every number $M > 0$ there is a corresponding number $\delta = \delta(M) > 0$, such that for all $(x, y) \in D$ with Euclidean distance $0 < \sqrt{(x-a)^2 + (y-b)^2} < \delta$, we have $f(x, y) > M$ (Bartle & Sherbert 2000; Stewart 2008).

The expanded expressions for true $g_1$, and true $g_2$ shown in Eq. (3.1a), and Eq. (3.1b) respectively, contain terms of the form $x/(x^2 + y^2)$, where $x$ and $y$ may represent $S_{11}$ or $S_{12}$. The expression for true $g_2$ given in Eq. (3.1b) can be considered without $R_{12}$ in the denominator to give terms of this form, as true $g_2$ is always multiplied by $R_{12}$ in Eq. (2.6) when calculating $b_{ij}$. With the multivariable infinite limit definition in Eq. (C1), it can be found that $\lim_{(x,y) \to (0^+,0)} x/(x^2 + y^2) = \infty$, and $\lim_{(x,y) \to (0^-,0)} x/(x^2 + y^2) = -\infty$. Therefore, as the $(x, y)$ variables approach $(0,0)$, the true $g_1$, and true $g_2$ expression terms tend towards negative infinity and infinity when the numerator is negative and positive, respectively. Hence, a discontinuity occurs where the numerator $x$ changes sign. The terms in the true $g_3$ expression in Eq. (3.1c) can be proven to tend towards infinity or negative infinity as $(x, y) \to (0, 0)$ with the same method.

## Appendix E: Smoothness of $b_{ij}$ Terms

The expanded expressions for true $g_1$, and true $g_2$ are given in Eq. (3.1a), and Eq. (3.1b), respectively. Substituting these into the 2D GEVH in Eq. (2.6) gives terms of the form $x^2/(x^2 + y^2)$ and $xy/(x^2 + y^2)$, where $x$ and $y$ may represent $S_{11}$ or $S_{12}$. To assess whether the limits that cause discontinuities in true $g_n$ terms also affect $b_{ij}$ terms, the same limits as those in **Appendix D** were taken on these functions. However, it was revealed that the limits do not exist at $(x, y) = (0,0)$. Nonetheless, by plotting $f(x, y) = x^2/(x^2 + y^2)$ and $g(x, y) = xy/(x^2 + y^2)$, it is understood that both functions are continuous and differentiable except where $(x, y) = (0,0)$, thereby demonstrating that discontinuities do not exist in these functions (Stewart 2008). Hence, any discontinuities found in true $g_1$ and true $g_2$ would be eliminated upon substitution in Eq. (2.6). In addition, $f(x, y)$ and $g(x, y)$ are bounded between 0 and 1, and between $-0.5$ and $0.5$, respectively. This implies that any changes in $f(x, y)$ or $g(x, y)$ between two points (which would be most significant near $(x, y) = (0,0)$) would not exceed these bounds. Substituting the true $g_3$ expression given in Eq. (3.1c) into Eq. (2.6) allows $(S_{11}^2 + S_{12}^2)$ to be cancelled out, thereby removing $S_{11}$ and $S_{12}$ from these $b_{ij}$ terms.

## Appendix F: Other Supplementary Input Variables

**Figure 18** shows contours of other supplementary input variables (SIV) for the periodic hills case that are commonly used for data-driven turbulence modelling.

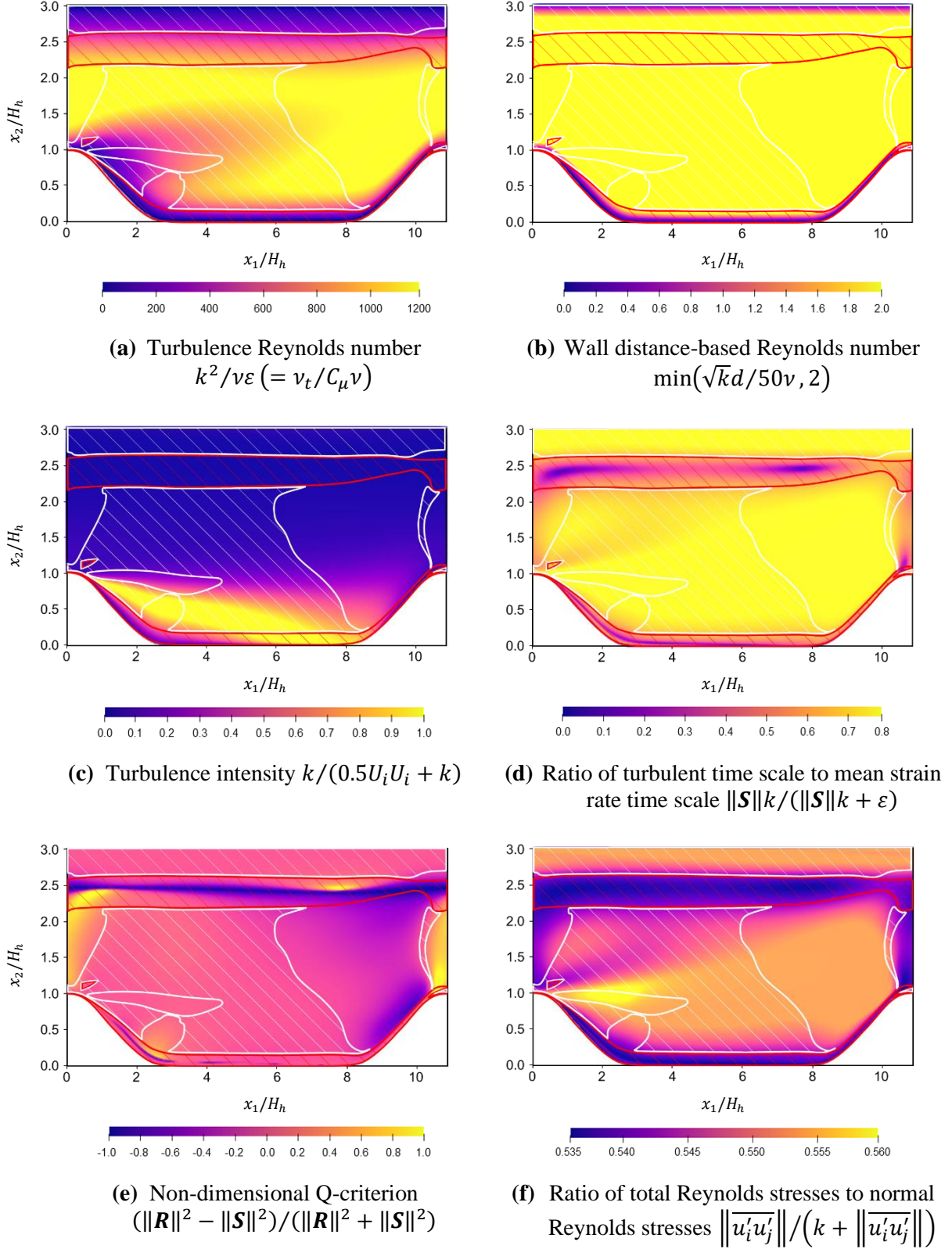

**(a)** Turbulence Reynolds number
$k^2/\nu\varepsilon \left(= \nu_t/C_\mu\nu\right)$

**(b)** Wall distance-based Reynolds number
$\min\left(\sqrt{k}d/50\nu, 2\right)$

**(c)** Turbulence intensity $k/(0.5U_iU_i + k)$

**(d)** Ratio of turbulent time scale to mean strain rate time scale $\|\boldsymbol{S}\|k/(\|\boldsymbol{S}\|k + \varepsilon)$

**(e)** Non-dimensional Q-criterion
$(\|\boldsymbol{R}\|^2 - \|\boldsymbol{S}\|^2)/(\|\boldsymbol{R}\|^2 + \|\boldsymbol{S}\|^2)$

**(f)** Ratio of total Reynolds stresses to normal
Reynolds stresses $\left\|\overline{u_i'u_j'}\right\|/\left(k + \left\|\overline{u_i'u_j'}\right\|\right)$

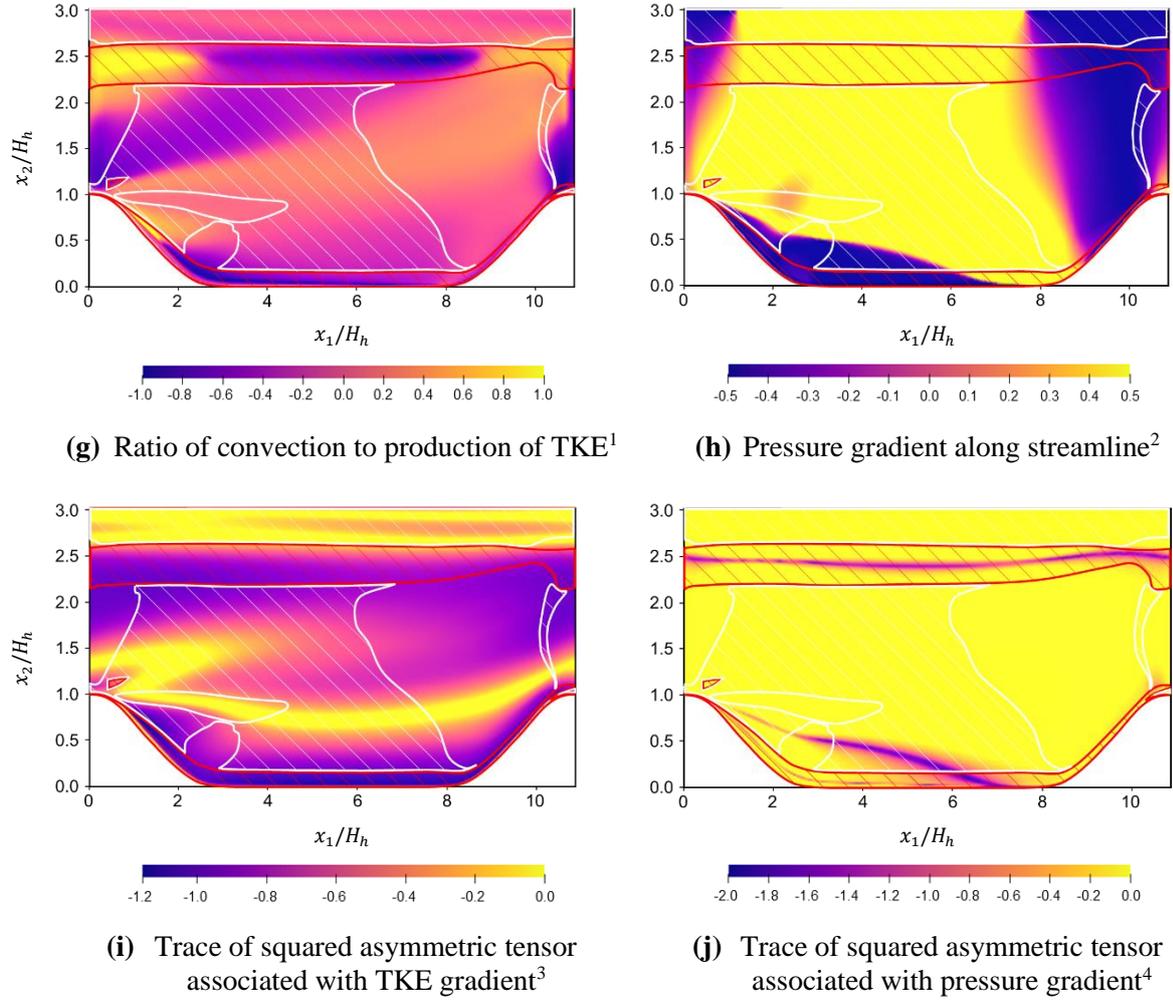

**(g)** Ratio of convection to production of TKE[1]

**(h)** Pressure gradient along streamline[2]

**(i)** Trace of squared asymmetric tensor associated with TKE gradient[3]

**(j)** Trace of squared asymmetric tensor associated with pressure gradient[4]

**Figure 18** Contour plots of commonly used supplementary input variables in data-driven turbulence modelling demonstrated with the periodic hills case. **(a)** turbulence Reynolds number, **(b)** wall distanced-based Reynolds number, **(c)** turbulence intensity, **(d)** ratio of turbulent time scale to mean strain rate time scale, **(e)** non-dimensional Q-criterion, **(f)** ratio of total Reynolds stresses to normal Reynolds stresses, **(g)** ratio of convection to production of TKE, **(h)** pressure gradient along streamline, **(i)** trace of squared asymmetric tensor associated with TKE gradient, and **(j)** trace of squared asymmetric tensor associated with pressure gradient.

---

[1] Ratio of convection to production of TKE $= \dfrac{U_i \frac{\partial k}{\partial x_i}}{\left|\overline{u_j' u_l'}\, S_{jl}\right| + \left|U_i \frac{\partial k}{\partial x_i}\right|}$

[2] Pressure gradient along streamline $= \dfrac{U_k \frac{\partial P}{\partial x_k}}{\sqrt{\frac{\partial P}{\partial x_j}\frac{\partial P}{\partial x_j}U_i U_i} + \left|U_l \frac{\partial P}{\partial x_l}\right|}$

[3] Trace of squared asymmetric tensor associated with TKE gradient, $tr\left(A_k^2\right) = -\dfrac{2k\left|\frac{\partial k}{\partial x_i}\right|^2}{\left(\sqrt{k}\left|\frac{\partial k}{\partial x_i}\right| + \varepsilon\right)^2}$

[4] Trace of squared asymmetric tensor associated with pressure gradient, $tr\left(A_p^2\right) = -\dfrac{2\left|\frac{\partial P}{\partial x_i}\right|^2}{\left(\left|\frac{\partial P}{\partial x_i}\right| + \left\|U_i \frac{\partial U_i}{\partial x_j}\right\|\right)^2}$